\documentclass[sigconf]{acmart}
\AtBeginDocument{%
  }

\copyrightyear{2026}
\acmYear{2026}
\setcopyright{cc}
\setcctype{by}
\acmConference[CHI '26]{Proceedings of the 2026 CHI Conference on Human Factors in Computing Systems}{April 13--17, 2026}{Barcelona, Spain}
\acmBooktitle{Proceedings of the 2026 CHI Conference on Human Factors in Computing Systems (CHI '26), April 13--17, 2026, Barcelona, Spain}
\acmPrice{}
\acmDOI{10.1145/3772318.3791655}
\acmISBN{979-8-4007-2278-3/2026/04}

\usepackage{subcaption}
\usepackage{enumitem}
\usepackage{geometry}
\usepackage{multirow}
\usepackage{colortbl}
\usepackage{hyperref}
\usepackage{xcolor}
\usepackage{tikz}
\usepackage[normalem]{ulem}
\usetikzlibrary{calc}

\newcommand{\newtext}[1]{\textcolor{black}{#1}}

\begin{document}

\title[Beyond Descriptions] {Beyond Descriptions: A Generative Scene2Audio Framework for Blind and Low-Vision Users to Experience Vista Landscapes}

\author{Chitralekha Gupta}

\affiliation{%
  \institution{Augmented Human Lab, \\Dept. of Computer Science, \\National University of Singapore}
  \country{}
}
\email{chitralekha@nus.edu.sg}
\orcid{0000-0003-1350-9095}

\author{Jing Peng}
\affiliation{%
  \institution{Augmented Human Lab, \\National University of Singapore}
  \country{}
  }
\email{peng_jing@u.nus.edu}
\orcid{0009-0002-4054-3547}

\author{Ashwin Ram}
\affiliation{%
  \institution{Saarland University, Saarland Informatics Campus}
  \city{Saarbrücken}
  \country{Germany}
}
\email{ashwinram10@gmail.com}
\orcid{0000-0003-1430-8770}

\author{Shreyas Sridhar}
\affiliation{%
 \institution{Augmented Human Lab, \\National University of Singapore}
 \country{}
 }
 \email{e0945817@u.nus.edu}
\orcid{0009-0004-6675-3459}

\author{Christophe Jouffrais}

\affiliation{%
  \institution{IRIT, CNRS, Toulouse, France}
  \country{}
  }
  \affiliation{%
  \institution{IPAL, CNRS, Singapore}
  \country{}
  }
  \email{christophe.jouffrais@cnrs.fr}
  \orcid{0000-0002-0768-1019}

\author{Suranga Nanayakkara}
\affiliation{%
  \institution{Augmented Human Lab, \\Dept. of Computer Science, \\National University of Singapore}
  \country{}
}
\email{suranga@ahlab.org}
\orcid{0000-0001-7441-5493}

\renewcommand{\shortauthors}{Gupta et al.}

\begin{abstract}
Current scene perception tools for Blind and Low Vision (BLV) individuals rely on spoken descriptions but lack engaging representations of visually pleasing distant environmental landscapes (Vista spaces). Our proposed Scene2Audio framework generates comprehensible and enjoyable nonverbal audio using generative models informed by psychoacoustics, and principles of scene audio composition. 
Through a user study with 11 BLV participants, we found that combining the Scene2Audio sounds with speech creates a better experience than speech alone, as the sound effects complement the speech making the scene easier to imagine. A mobile app “in-the-wild” study with 7 BLV users for more than a week further showed the potential of Scene2Audio in enhancing outdoor scene experiences. Our work bridges the gap between visual and auditory scene perception by moving beyond purely descriptive aids, addressing the aesthetic needs of BLV users.
\end{abstract}


\begin{teaserfigure}
    \includegraphics[width=\linewidth]{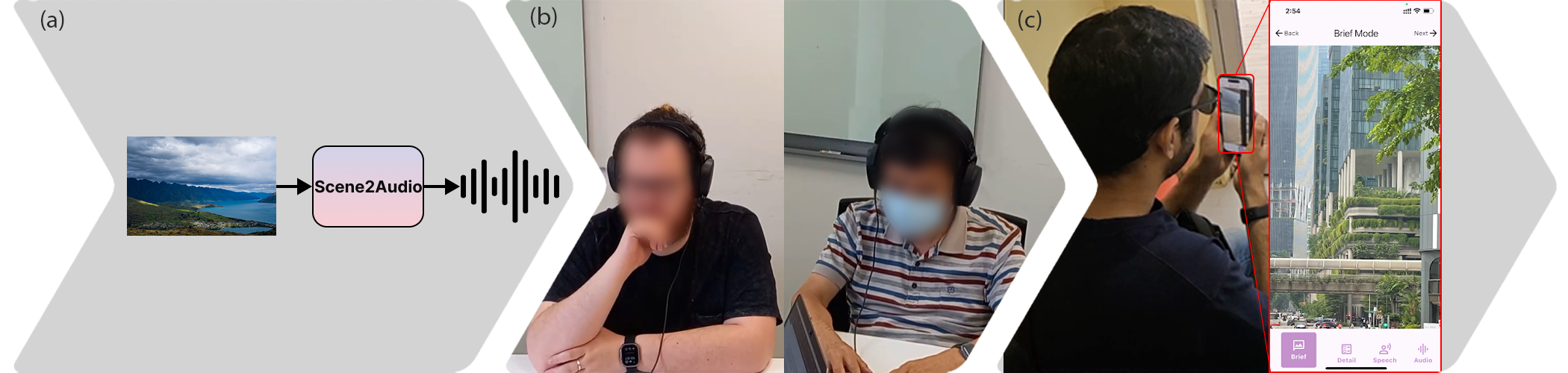}
  \caption{This paper consists of three stages: (a) Designing and evaluating the Scene2Audio Framework, (b) A lab-study with BLV participants to evaluate the sounds generated from Scene2Audio framework, and (c) In-the-wild evaluation of user experience of this framework through a mobile app user study with BLV participants for over a week.%
 }
  \Description{This image is a flowchart illustrating the three-stage research process described in the paper. The stages are arranged horizontally from left to right, connected by arrows indicating the progression of the work.
From left to right:
(a) Stage 1 - Framework Development: This stage is represented by a block diagram representing a scenic image passing through the Scene2Audio framework to generate an audio. 
A right-facing arrow connects Stage 1 to Stage 2.
(b) Stage 2 - Lab Study: This stage is represented by photos of two BLV participants sitting at a computer within a controlled environment, wearing headphones. This symbolizes the controlled lab study with Blind and Low-Vision (BLV) participants to evaluate the quality and effectiveness of the sounds generated by the framework.
Another right-facing arrow connects Stage 2 to Stage 3.
(c) Stage 3 - Field Study: This stage is represented by the photo of a BLV person holding a mobile phone, and a zoomed in picture of the mobile phone screen is on the right that contains the scene2audio app called sonic vista with a vista scene image and four audio icons at the bottom - Brief, Detail, Speech, Audio. This symbolizes the in-the-wild user study where BLV participants used the framework via a mobile app in their daily lives for over a week to evaluate the real-world usage of the system.}
  \label{fig:teaser}
\end{teaserfigure}


\maketitle

\section{Introduction}
According to the classification of psychological spaces provided by Montello et al.~\cite{montello1993scale}, a Vista is the far-field space that can be visually apprehended from a single place, without appreciable locomotion, for example the horizon. 
For blind and low-vision (BLV)\footnote{We understand it is important to use neutral and accurate language to describe people with disabilities, as it promotes respect and inclusivity. Following the recommendations from the SIGACCESS community \cite{hanson2015writing}, we use the term \textit{``Blind and Low Vision (BLV)''}, when referring to individuals with vision loss. The use of this abbreviation is meant for ease of reference and consistent with prior work~\cite{gonzalez2024investigating,jiang2023beyond}. We also acknowledge that other terms might also be appropriate \cite{sharif2022should}, such as \textit{``People with Visual Impairments (PVI)''}} users, experiencing distant scenic views, such as city skylines or mountain vistas, remains a significant accessibility challenge. These Vista spaces ~\cite{montello1993scale} play a crucial role not only in orientation but also in leisure, emotional well-being, and cognitive engagement \cite{maslow1975motivation, mcleod2007maslow, gupta2024imwut}.  Existing technologies can provide verbal descriptions for scene understanding, but they often feel transactional, imposing cognitive load, and failing to evoke the aesthetic pleasure of sighted viewing \cite{ji2021seeing,dubus2013systematic}. 

\begin{figure*}
    \centering
\includegraphics[width=\linewidth]{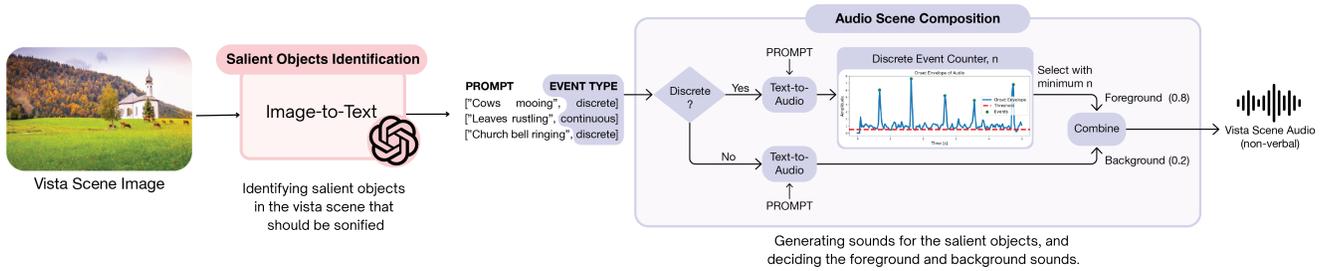}
    \vspace{-0.8cm}
    \caption{Overview of Scene2Audio Framework.}
    \Description{The figure illustrates a pipeline for generating non-verbal audio from a vista scene image.
On the left, there is a photo of a grassy field with cows, autumn trees, and a church building. Below it, the label reads “Vista Scene Image.”
An arrow points to a pink box titled “Salient Objects Identification.” Inside it, the text reads “Image-to-Text” with a ChatGPT icon. Beneath the box, there is a caption: “Identifying salient objects in the vista scene that should be sonified.”
From this box, arrows lead to the right. A list is shown:
[“Cows mooing”, discrete]
[“Leaves rustling”, continuous]
[“Church bell ringing”, discrete]
The column headers are PROMPT and EVENT TYPE.
The process continues to a section labeled “Audio Scene Composition” (outlined in purple).
A diamond-shaped decision box asks: “Discrete?”
If Yes, the arrow leads to a Text-to-Audio box. The output goes into a panel labeled “Discrete Event Counter, n” showing a line graph with peaks labeled “Onset Envelope of Audio” and “Threshold Events.” The text says: “Select with minimum n.” This path continues to a label “Foreground (0.8).”
If No, the arrow leads to another Text-to-Audio box, labeled PROMPT, which continues to a label “Background (0.2).”
Both foreground and background paths feed into a Combine box.
The final output points to a soundwave icon with the label “Vista Scene Audio (non-verbal).”
At the bottom of the purple section, a caption reads: “Generating sounds for the salient objects, and deciding the foreground and background sounds.”
}
    \label{fig:overview}
    \vspace{-0.2cm}
\end{figure*}

\textit{How can we design an assistive framework that enables BLV users to not only understand but also enjoy Vista spaces through audio, while minimizing cognitive overhead?} In this paper, our main aim was to design a system that can automatically generate non-verbal sounds corresponding to Vista spaces, which provides scene understanding as well as an enhanced experience to BLV. 
For this, we propose the ``\textit{Scene2Audio}'' framework, which has a two-step process (Figure \ref{fig:overview}). In the first step, it generates non-verbal representative audio for the different objects in the scene (\textit{Salient Objects Identification}). In the second step, it combines those individual audio components to create a representative audio for the scene, informed by psycho-acoustics, auditory scene analysis, and Foley sound synthesis \cite{moore2012introduction,bregman1994auditory,chion2019audio} (\textit{Audio Scene Composition}). Through a listening test with 21 sighted participants, we evaluated the sounds generated from \textit{Scene2Audio} framework in terms of comprehending the vista image. We found that the sounds from our framework significantly outperform those generated with existing baseline techniques.

We then investigated different strategies to use our \textit{Scene2Audio} framework in combination with verbal descriptions to provide an engaging experience of a distant scene for BLV. We conducted a user study with 11 BLV participants to evaluate these strategies, and found that overlaying verbal description with representative non-verbal sounds provide a better user experience than verbal description of the scene only. 
Finally, we conducted a week-long in-the-wild study with 7 BLV participants using our framework integrated in a mobile smartphone app. The participants used our app in their own time to click images and gave feedback through the app for scenes they were keen to experience. During this time, a total of 70 scene images were clicked by the participants and feedback about their experience of the different audio modes was obtained. This final study investigated the ecological validity of sound effects generated from Scene2Audio and showed its potential in real-world contexts. 

Figure \ref{fig:teaser} summarizes the three stages of this study. Specifically, our contributions in this paper are three-fold:
\begin{itemize}
    \item We introduce Scene2Audio, a framework that generates non-verbal soundscapes for Vista space scenes by combining AI-driven object sonification with psychoacoustic scene composition. 
    \item Through a lab-based user study with 11 BLV participants, we identify sound rendering strategies for combining verbal descriptions with the generated non-verbal audio to improve the experience of Vista space. 
    \item An in-the-wild deployment of the system for more than a week via a mobile app reveals how BLV users (N=7) can use this system in their daily life, uncovering opportunities for leisure-oriented assistive technologies.
\end{itemize}

\section{Background \& Related Work}
\subsection{Non-Verbal Audio for Mental Representation of the Scene}
In audio narratives such as radio, podcasts, and audiobooks, sound effects compensate for the absence of visuals by supporting comprehension, guiding mental imagery, and enhancing attention and recall~\cite{rodero2020audio, bolls2003saw, steinhaeusser2021comparing, rodero2012see, rodero2019spark, potter2006made}. In games and films, they complement visuals to advance narrative, foster immersion, and convey realism~\cite{murray2019sound, rodero2020audio}.   Sound effects typically serve three functions: reinforcing actions, simulating atmosphere, and evoking emotions~\cite{green2000role, steinhaeusser2021comparing}. Their non-verbal use is therefore critical for scene awareness when images are absent, motivating our approach to automatically generating such effects for describing vista spaces, aimed at enhancing scene experience for BLVs.

As Crisell~\cite{crisell1994understanding} and Fryer \cite{fryer2010audio} stated, sound effects, when used effectively, can stimulate our imagination. However, to convey clear meaning, verbal description is necessary along with the sound effects. In the context of enhanced audio descriptors (EAD) for movies designed for BLVs, Lopez et al.~\cite{lopez2021enhancing} and Jiang et al.~\cite{jiang2023beyond} showed that audio effects complement verbal descriptions and reduce cognitive load. And Hattich et al.~\cite{hattich2020hear} showed that such audio effects improve immersion. Furthermore, combining audio with speech during storytelling has shown greater induction of emotions and transportation to the scene compared to speech alone among people with BLV conditions \cite{steinhaeusser2021comparing}. Therefore, we designed strategies to combine verbal descriptions with AI-generated audio effects, and validated their effectiveness in providing an enhanced user experience. 

\subsection{Non-Verbal Audio in Assistive Devices for BLV Beyond Navigation}
In recent HCI work, there is a growing interest in assistive devices for BLVs that go beyond functional aids to enhancing cognitive and aesthetic experiences, such as providing accessible media content \cite{liu2021makes,seo2018understanding,mcdonnell2024caption}, sports content \cite{jain2023towards}, visual arts \cite{li2023understanding}, and mixed reality experiences \cite{may2020spotlights,bandukda2020audio,zhao2019seeingvr} through verbal and non-verbal sounds. Previous research has also emphasized the importance of the experience of outdoor nature scenes in fulfilling these cognitive and aesthetic needs in people with BLV conditions \cite{gupta2024imwut,bandukda2020places,bandukda2019understanding}. Experiencing the Vista Space involves not only communicating information about the objects in the scene, but also providing an engaging experience. 

Non-verbal audio, such as auditory icons, is often used in assistive applications for BLVs for object awareness, i.e.~identification and localization~\cite{constantinescu2020bring,tomlinson2020auditory}. Auditory icons are brief sounds that have semantic connections to the objects, functions, and actions they represent. They take advantage of the user's prior knowledge regarding auditory associations between sounds and their meaning, such as the sound of a siren to indicate an emergency or the sound of a crinkling paper to denote the move of a file to the trash can on a user interface~\cite{gaver1987auditory}. However, synthetically designing auditory icons requires parametric sound modeling, which involves designing sounds for individual objects. This process lacks scalability due to the manual crafting of model parameters, which demands expert knowledge. This becomes more challenging when a scene needs to be sonified, as it may consist of multiple objects simultaneously present in the field of view. An alternative approach might involve retrieving sounds from existing audio data sets (e.g.,~\cite{gemmeke2017audio}). However, even large sound libraries may lack specific audio samples for certain contexts, and they offer little flexibility in dynamically adjusting acoustic properties to suit different scenarios. Generative models present a promising solution by learning a flexible sound space that can synthesize auditory icons on demand. Unlike static data sets or manual design, this approach balances naturalness with adaptability, enabling scalable sound generation for diverse real-world contexts.

Gupta et al.~\cite{gupta2024imwut} found that the sounds for specific objects generated from current AI-based audio generative models are comparable in quality and intuitiveness to the auditory icons designed for those objects by experts. However, there is a significant degradation in the audio quality of AI-generated sounds for scenes with multiple objects or events compared to those for scenes with only one object or event. Although AI models excel at generating sounds for single objects, they struggle with scene-level composition. They are predominantly trained with data sets consisting of objects and event labels \cite{liu2023audioldm,sheffer2023hear}, rather than on scene soundscapes that include multiple objects or events occurring simultaneously. Our framework bridges this gap by decomposing scene images into salient objects and recomposing the auditory representation of the scene using sound synthesis principles.


\subsection{Generative Audio Techniques and Limitations}

Designing images \cite{sheffer2023hear}, and text \cite{liu2023audioldm} to representative nonverbal audio is an active area of research. 
The state-of-the-art image-to-audio (nonverbal) model, Im2Wav \cite{sheffer2023hear}, consists of a transformer-based audio generative model that is conditioned on image representations obtained from a pre-trained CLIP model \cite{radford2021learning}. The DCASE 2023 Challenge (Task 7 Foley Sound Synthesis)\footnote{\url{https://dcase.community/challenge2023/task-foley-sound-synthesis-results}} showed the viability of AI-generated nonverbal sound effects, such as footsteps or rain, using models conditioned on sound categories. Text-to-audio (nonverbal) models such as AudioGen \cite{kreuk2022audiogen} and AudioLDM \cite{liu2023audioldm} are capable of generating high-quality audio from textual input, beyond categories. For example, AudioGen is a state-of-the-art autoregressive, transformer-based audio generation model that generates audio from text description by using textual features as a conditioning signal. Recent work has demonstrated the use of such AI models for the application of generating sounds for individual objects in an augmented reality experience \cite{su2024sonifyar}. 

Existing mobile applications assisting BLVs to be aware of their surroundings are typically image-to-text-to-speech frameworks. For example, SeeingAI\footnote{\url{https://www.seeingai.com/}} and BeMyAI\footnote{\url{https://www.bemyeyes.com/blog/introducing-be-my-eyes-virtual-volunteer}} in the BeMyEyes app use state-of-the-art image-to-text models like OpenAI's ChatGPT-4\footnote{\url{https://openai.com/index/gpt-4/}} to provide textual descriptions of images. These descriptions are then converted to speech using text-to-speech accessibility features such as VoiceOver\footnote{\url{https://support.apple.com/en-sg/guide/voiceover/welcome/mac}} on iPhones and TalkBack\footnote{\url{https://support.google.com/accessibility/android/answer/6283677?hl=en}} on Android devices. Gonzalez et al.~\cite{gonzalez2024investigating} investigated the use cases and motivations of BLVs when using ``scene description'' or image-to-text apps through a two-week diary study. They found that one of the top three goals of using the scene description app is ``building understanding of the scenery''. Although speech provides direct and rich information, it often results in a high cognitive load and can interfere with natural voice and sound inputs~\cite{ji2021seeing,dubus2013systematic}. Thus, in this work, we explored alternative sonification methods for understanding as well as enjoying Vista Spaces.

\section{Implementation of Scene2Audio framework for Vista Space Sonification}
\label{sec:proposedfw}
\subsection{Overview}
According to the classification of psychological spaces provided by Montello et al.\cite{montello1993scale}, a Vista is the far-field space that can be visually apprehended from a single place, without appreciable locomotion, for example the horizon. In this work, we consider these distant environmental scenes in the Vista space. These scenes although are within visible range, are beyond the audible range. Moreover, actions occurring in this far field are not expected to be prominent. Hence, we consider a static image of the scene to be a closer approximation of a distant environmental scene, or a Vista space scene.

Our goal is to convey the key elements and the general atmosphere, or \textit{ ``gist''}, of a Vista space scene through non-verbal audio \cite{oliva2001modeling,harding2007auditory}. A straightforward approach would be to use a state-of-the-art generative model such as Im2Wav \cite{sheffer2023hear} to convert an image to an audio clip. As previously mentioned, such models are typically trained for one object or event in focus. Vista spaces, on the other hand, often have several objects in the scene, which is difficult to sonify using existing image-to-audio models. Moreover, training a model on scene data is difficult due to a lack of existing datasets that consist of images of complex Vista spaces and the corresponding audio that conveys its gist. 

Alternatively, one could create a pipeline consisting of an image-to-text followed by a text-to-audio model, where the image-to-text model describes the scene and the text-to-audio model generates the audio corresponding to the description. However, existing text-to-audio models are also trained on individual objects and events and are not designed to handle descriptions of a complex Vista space. Indeed, a listening test evaluation (see Section \ref{sec:proposed_eval}) of Im2Wav and image-to-text-to-audio models showed that the resulting audio would tend to capture only one of the objects in the description. 





To design our Scene2Audio framework, we applied the `principle of designing for mental models' recommended as a design principle for generative AI applications \cite{weisz2024design}, ensuring that the generated auditory scenes are consistent with how the human auditory system processes and comprehends sound.
Our proposed Scene2Audio framework includes two components: (1) Salient Objects Identification: to identify key sound-making or sonic objects within the scene and describe them by action phrases (noun+verb) using an image-to-text model, and (2) Audio Scene Composition: to generate sounds for each object, layered to create a balanced and immersive auditory experience. We use the prompt chaining technique\footnote{\url{https://www.promptingguide.ai/techniques/prompt_chaining}} across the two AI models (image-to-text and text-to-audio). The image-to-text model is prompted with a subtask, and then its response is used as an input prompt for the text-to-audio model. 
Figure \ref{fig:overview} provides the details of the two components of the proposed framework. The following subsections detail the design of this framework.

\subsection{Salient Objects Identification}
\newtext{Although not all visually present objects produce sound, our design follows the ecological principle that humans build auditory mental models using sonic entities \cite{gaver1993world, bregman1994auditory}.
Vista scenes contain many visually salient but acoustically silent objects (e.g., buildings, mountains). Sonifying these would violate ecological expectation and overload listeners.
Thus, our system focuses specifically on “sonic objects”, i.e.~elements that plausibly produce environmentally meaningful sounds. Sonic objects would include animate sound-making elements (e.g. birds, dogs) as well as contextual sound-making events in a scene (e.g. wind blowing, leaves rustling). This selective mapping better matches how BLV individuals use sound to infer scene structure, and avoids constructing an artificial “all-objects-have-sound” mental model.}

The human auditory system naturally separates complex environments into distinct sound events and objects \cite{bregman1994auditory, moore2012introduction}. 
By sonifying these sonic objects separately, we mimic this ability, ensuring that each sonic element retains its distinct identity. Additionally, focusing solely on sonic objects, rather than the entire scene, leverages the strengths of text-to-audio models, which are trained on naturally occurring sounds. This approach should make the generated audio more intuitive and hence easier to comprehend. 

To extract sonic objects from the scene image, we use the following prompt with GPT-4: \textit{``Describe this image with respect to the sound-making objects in the scene.''}. 
Once the sonic objects are identified, generating appropriate sounds for these objects requires a carefully designed set of steps as follows: 

\subsubsection{Generate Action Phrases.} A fundamental difference between visual perception and auditory perception is how we perceive objects.
We see visual objects, but we hear \textit{sound events} \cite{murray2019sound, gaver1987auditory}. Although sounds may originate from an object (such as a person or a vehicle), they are triggered when an event on or by the object occurs. For example, leaves (the object) produce sound when they are moved by the wind (the event). Thus, we associate each identified sonic object with an action phrase, i.e.~a noun paired with a verb. The noun represents the object, and the verb represents the action that produces the sound. 
We prompt GPT-4 with \textit{``Only provide a noun+verb like phrase such that the noun is the sound making object and the verb is the possible action on or by the object that could create a sound''}.

\subsubsection{Identify the Sound Event Type.} The human auditory system processes different types of sound events, continuous and discrete, differently \cite{moore2012introduction}. Continuous sounds, such as wind or ambient noise, typically form the background of an auditory scene and are perceived as less salient. In contrast, discrete sounds, such as a bell ring or a door slam, are more likely to capture attention due to their distinct temporal boundaries.

\newtext{In this work, we adopt the standard auditory-perception distinction: (i) discrete sounds contain isolated acoustic events with clearly identifiable onsets/offsets, such as footsteps, knocks, or animal calls, and (ii) continuous sounds have sustained temporal energy without salient event boundaries, such as wind, waves, or running water \cite{bregman1994auditory}. Because these categories directly correspond to how the auditory system organizes environmental scenes, we use them as labels in the Audio Scene Composition stage.} 
To provide the Audio Scene Composition component with the event-type label, we prompt GPT-4: “At the end of each noun+verb phrase, attach a label `discrete' or `continuous' depending on whether the sound produced by the object is generally discrete or continuous.”

\subsubsection{Prompt Verification and Consistency Experiments}: \\
\noindent\newtext{{\textbf{Verification of Discrete and Continuous Labels: }}}
\newtext{
To ensure that the image-to-text model applied these definitions consistently, we conducted an offline quantitative evaluation on 100 images from Open Images Dataset \cite{OpenImages}, each containing multiple sound-making objects. For each GPT output, we extracted the noun in the noun+verb phrase (e.g., \emph{wind blows}, \emph{footsteps echo}), and assigned a ground-truth label (discrete or continuous) using a hybrid protocol. First, we relied on a category-level mapping for sound sources whose acoustic character is widely established in auditory perception (e.g., \emph{bird}, \emph{dog}, \emph{footsteps} as discrete; \emph{wind}, \emph{waves}, \emph{river} as continuous). Second, for visually identified sources whose acoustic profile depends on scene context (e.g., \emph{people}, \emph{leaves}, \emph{wood}), one author applied the same operational definitions to assign labels. This annotation was performed carefully and consistently using the same criteria used to design the framework, and served as a targeted sanity check rather than a subjective coding exercise.}

\newtext{In total, we obtained ground-truth labels for 141 sonic objects across the 100 images. GPT’s event-type predictions achieved an accuracy of 0.96 (Cohen’s $\kappa = 0.91$), indicating high agreement 
between the model and our definitions of discrete and continuous sound sources. In reliability analysis, $\kappa$ values above 0.81 are widely interpreted as ``almost perfect agreement'' between raters \cite{landis1977measurement,mchugh2012interrater}, here reflecting alignment between the prompt design and the model’s internal categorization. The remaining disagreement arose from visually identified categories whose acoustic character varies with context (e.g., people talk, wood creaks). 
These results suggest that, for the types of sonic objects likely to be present in vista scenes, GPT's discrete/continuous labelling is largely consistent with our intended technical definitions. }

\newtext{In addition, for sounds labeled as discrete, we applied a secondary signal-processing validation step: generating multiple audio candidates and selecting the one with the lowest number of detected amplitude peaks ($n\geq1$) (See Section \ref{sec:mindiscrete}). This two-stage process ensures that the “discrete” label results in perceptually appropriate discrete-event audio, even when text-to-audio models produce variable outputs.}

\noindent\newtext{\textbf{Consistency test for Discrete and Continuous Labels:} } 
\newtext{To evaluate stability of the image-to-text model, we ran five independent passes on the 100 images from Open Images Dataset with identical prompts, as mentioned above. Across these robustness tests, the event-type showed an agreement (measured as the average of ratio of the number of most common event-type identified over the number of runs) of an average of $0.79\pm0.18$ across runs over the 100 images, indicating that event-type assignments are moderately stable. Following standard interpretations of categorical agreement, values in the range 0.61–0.80 are commonly described as reflecting “substantial agreement” between raters \cite{landis1977measurement}, suggesting that the prompt-based assignment is within an accepted reliability range. This indicates a favorable stability of the event label from the image-to-text model for the downstream audio generation pipeline.}\\
\newtext{\textbf{Consistency test for Action phrase generation:}} \newtext{To evaluate the robustness of the noun–verb action phrases generated by our image-to-text module, we conducted a multi-run consistency test. For each image in the dataset, we queried the model five times using the same prompt and image, extracted the set of generated noun+verb phrases, and computed a phrase-level cosine similarity metric. For every phrase in each run, we identified its closest semantic match among phrases produced in all other runs, measured via SentenceTransformer embeddings (MiniLM-L6-v2), and averaged these maximal similarities across all phrases and runs. This metric captures semantic stability even when phrasing varies slightly (e.g., “waves crashing” vs. “water splashing”). Across all images, the model achieved an average phrase-level cosine similarity of $0.92\pm0.054$, indicating that the noun–verb action phrases remain highly consistent across repeated generations of the same scene.}

\subsection{Audio Scene Composition}
In distant scenes, the exact spatial positioning of objects may be less critical to understanding the overall auditory representation of the scene \cite{bregman1994auditory,gaver1993world}. Therefore, in the Audio Scene Composition component of our framework, we assume that all objects in the Vista space are equidistant from the user's point of view. \newtext{Our primary goal in this study was to isolate and evaluate the contribution of the content generated by the Scene-to-Audio framework, rather than the contribution of spatialization or distance rendering. Spatialization is well known to substantially influence immersion \cite{potter2022relative}; however, incorporating it at this stage would introduce an additional perceptual factor that could confound our comparisons across conditions. For example, a speech-only audio cannot be spatialized in a comparable or meaningful way, making it difficult to attribute differences in comprehension or immersion to the framework’s generative audio rather than to spatial rendering itself. By holding spatial distance constant, we ensured a controlled and methodologically fair comparison between conditions, allowing us to evaluate whether the generated non-verbal audio meaningfully improved comprehension and experience on its own.}

Figure \ref{fig:overview} presents details of this component, and the key decisions are explained below.

\subsubsection{Differentiating Discrete and Continuous Sounds}
We treat discrete and continuous sounds differently, as this distinction is crucial to achieve a realistic and immersive auditory experience, grounded in psychoacoustics and auditory scene analysis \cite{moore2012introduction,bregman1994auditory}. Discrete sounds, such as footsteps or a car horn, are characterized by isolated audio events with clear onsets and offsets. These sounds naturally draw attention and are perceived to be in the foreground, regardless of their actual distance from the listener. In contrast, continuous sounds, such as the hum of a distant highway or the rustling of leaves, create a persistent background layer. These are generally not the main focal point of the scene and are perceived as ambient elements, contributing to the overall atmosphere without dominating it. Thus, we inform the audio scene composition component with the event-type label (i.e.~discrete or continuous) generated in the previous step.

\subsubsection{Minimizing Discrete Sound Events}
\label{sec:mindiscrete}

Psychoacoustic research indicates that repetitive, high-intensity discrete sounds can cause discomfort and auditory fatigue \cite{kryter2013effects}. Continuous exposure to such sounds can result in auditory masking \cite{moore2012introduction}, where these prominent sounds overshadow other auditory cues, reducing the clarity and immersiveness of the experience. To avoid these issues, we incorporated a mechanism to minimize the number of events in the generated discrete sounds. 

In current text-to-audio models, there is no straightforward way to control the number of discrete events in the generated sound, as these models are trained on audio with captions that do not elicit this granularity in semantics \cite{kamath2024morphfader,xie2024picoaudio,kamath2024example, gupta2023towards}. For example, a typical text label would be ``a stick hitting a wooden table'', but it would not mention the frequency of hits or the number of hits. Thus, attempts to use different text prompts to manage the number of events yield inconsistent results with these models \cite{kamath2024example}. Therefore, we adopted a signal processing-based method to achieve the desired control. Specifically, when a text prompt is tagged as ``discrete'', we generate ten audio clips, each five seconds long, using the text-to-audio model (AudioGen). We then analyze each audio clip using peak detection in the amplitude envelope. 
\newtext{We compute the onset envelope using \texttt{librosa.onset.onset\_strength} (librosa 0.11.0) with default parameters. Event detection is performed using \texttt{librosa.onset.onset\_detect} that uses a peak picking algorithm where the parameters are chosen by large-scale hyper-parameter optimization\footnote{\url{https://github.com/CPJKU/onset_db}} with \texttt{hop\_length = 512} samples. The sampling rate of the audio is 16 kHz.
} 
This method allows us to count the number of discrete events within each clip and select the one with the least number of events $n$ ($n\geq1$). This increases the likelihood that the chosen sound is not overly repetitive or jarring, thereby providing a more comfortable listening experience.

\subsubsection{Foreground and Background Layering}

To create a balanced auditory scene, we performed a weighted average of all the sounds, assigning different weights to discrete and continuous sounds. \newtext{Before mixing, each generated audio stem is peak-normalized using \texttt{librosa.util.normalize} to remove level differences introduced by the text-to-audio model. We then apply a simple weighted-sum mix: discrete sounds at 0.8 and continuous sounds at 0.2. This choice was motivated by Foley mixing conventions \cite{beauchamp2012designing,chion2019audio}, where foreground events must remain perceptually salient while background ambience provides spatial context without overpowering the listener.}

By placing discrete sounds in the foreground and continuous sounds in the background, we mimic how the human auditory system prioritizes and segregates sound sources \cite{moore2012introduction,bregman1994auditory}. This approach ensures that the auditory scene remains coherent and does not overwhelm the listener with too many competing elements.


\newtext{<Removed the Comparison of the framework with baselines>}

\section{Using Scene2Audio Framework for Enhancing Scene Experience of BLV}
\newtext{A short evaluation of the framework with sighted participants (Appendix \ref{sec:proposed_eval})} showed that although nonverbal audio generated from the Scene2Audio framework can create a comprehensible and pleasant experience, these sounds alone can sometimes lead to ambiguity, as the sounds may have multiple interpretations. For example, a splash of water could represent a water body, but how big the water body is, would be unclear if one listens to that sound alone. However, verbal descriptions, while clear, often lack engagement. Thus, we explore strategies to combine non-verbal audio with verbal descriptions for a more engaging and immersive experience.

We investigate the impact of auditory feedback variants that combine non-verbal and verbal sounds, aiming to enhance scene comprehension and user experience. Such combinations of nonverbal and verbal sounds have previously been explored in the context of making art exploration accessible \cite{rector2017eyes,morris2018rich}. However, to the best of our knowledge, these strategies have not been investigated in the context of Vista space experience. 
We designed four strategies of audio feedback\footnote{Readers are encouraged to listen to these four Audio Types for all the scenes on our webpage: \url{https://scenetoaudio.github.io/scenetoaudio/\#/section-4}} for a given Vista space, which we call \textit{Audio types}:

\begin{itemize}
    
\item \textbf{Speech-only}: Speech or spoken description is considered as the baseline approach that represents the auditory feedback mechanism of existing apps for scene understanding (eg. SeeingAI and BeMyAI) used by BLVs. It consists of a short spoken description of the scene.
\item \textbf{Audio-only}: This is the non-verbal representative audio of a scene generated by our Scene2Audio framework. 
\item \textbf{Overlay}: The verbal description (same as Speech-only condition) is overlaid on the non-verbal sound (same as Audio-only condition) of the scene, with the verbal description in the foreground.
\item \textbf{Overlay-Concat}: 
Sonifying different sonic objects in the scene and overlaying each one with the corresponding speech description. They were then played sequentially. This included more details about the scene along with the relative positions of the different sonic objects in the scene with respect to the user. 
For example, for the Vista space image in Figure \ref{fig:overview}, it would generate the following overlay of text-to-speech and scene-to-audio: \textit{``A group of cows is directly ahead in a lush, green meadow.''}; followed by \textit{``To your right, trees with leaves in autumn hues are on a slope.''};  followed by \textit{``A small white church with a steeple sits in the distance, slightly to the left.''}. 
\end{itemize}

\newtext{It is to be noted that we did not include human-written scene descriptions as a baseline because they are not scalable for real-world use. Current BLV tools (e.g., SeeingAI, BeMyAI, Envision) rely on LLM-generated descriptions, not human-authored ones, since human captions cannot be produced consistently or on demand for arbitrary scenes. Thus, using LLM-generated descriptions therefore provides a methodologically consistent and a deployable baseline for our use case.}

\noindent\newtext{\textbf{Consistency test for the short description generation prompt:} } 
\newtext{To evaluate stability of the image-to-text model for short description generation, we ran five independent passes on the 100 images from Open Images Dataset \cite{OpenImages} with identical prompts, i.e.~``\textit{Provide a short sentence (<10 words) describing the scene to a blind person}'', and found an average cosine similarity of $0.84\pm0.11$ between the embeddings of the short textual description generated across the runs over the 100 images, showing reasonable stability of the outputs across runs.}

In a lab study, as discussed in the following sub-sections, we evaluated the impact of each of these strategies on the experience of the Vista space for a set of preselected scenes. 

 
\subsection{Participants}
The study involved 11 BLV participants, including 9 males and 2 females, ranging from 27 to 73 years of age (mean age 49.8 years, std. dev 17.1). The participants had a variety of visual conditions, such as Retinitis Pigmentosa, Glaucoma, and Macular Degeneration. Most of the participants developed visual impairments after the age of 14, with the exception of one participant (P4) who has been visually impaired since birth. Visual acuity varied between participants, with most of them reporting a significant vision loss (e.g., less than 2\% to 10\% of residual vision). Most of the participants were familiar with AI-based image-to-text apps like BeMyEyes, Seeing AI, and Google Lookout, which helped them understand the additional sound effects and engage with the experiment. Detailed demographics of the participants are in Table \ref{tab:pvi_details}. The participants were recruited through local organizations for the visually impaired. We obtained ethics approval from the Institute Review Board (IRB) before conducting the user studies. 

\begin{table*}[]
\centering
\caption{Details of the BLV participants}
\label{tab:pvi_details}
\resizebox{0.98\textwidth}{!}{%
\begin{tabular}{
    >{\centering\arraybackslash}p{0.5cm} 
    >{\centering\arraybackslash}p{0.8cm} 
    >{\centering\arraybackslash}p{0.9cm} 
    >{\arraybackslash}p{3.2cm} 
    >{\centering\arraybackslash}p{1.5cm} 
    >{\centering\arraybackslash}p{1.5cm} 
    >{\arraybackslash}p{2.5cm} 
    >{\arraybackslash}p{5cm}
}
\rowcolor{lightgray!80}
\hline
\textbf{ID} & \textbf{Age} & \textbf{Gender} & \textbf{Visual Status/ Condition} & \textbf{Visual Acuity} & \textbf{Onset Age} & \textbf{Type of Earphones used} & \textbf{Apps to Know About Surroundings} \\ 
\hline
\rowcolor{lightgray!40}
1 & 30 & M & Legally Blind & <2\% & 25 & Wired and wireless earphones & BeMyEyes, Seeing AI \\ 
\rowcolor{white}
2 & 73 & M & Retinitis Pigmentosa - \textit{``I can differentiate between light and shadow''} & not known & 59 & Bone conduction headphones (open ear) & Moovit - public transport navigation, BeMyEyes, BeMyAI, Lazarillo \\ 
\rowcolor{lightgray!40}
3 & 45 & M & Legally Blind & <2\% & 16 & Over the ear headphones & Seeing AI \\ 
\rowcolor{white}
4 & 27 & M & Retinoblastoma, \textit{``I can still see colors and shapes, but cannot see details''} & 20/200 & Since birth & AirPods & None \\ 
\rowcolor{lightgray!40}
5 & 60 & M & Retinitis Pigmentosa - a genetic disorder & <10\% & 41 & Wired earphones & None \\ 
\rowcolor{white}
6 & 35 & M & Legally Blind & <2\% & 19 & In-ear headphones & BeMyEyes, BeMyAI \\ 
\rowcolor{lightgray!40}
7 & 51 & M & Retinitis Pigmentosa (RP), genetic - \textit{``I can only perceive shadow and light''} & <5\% & 30 & Bluetooth earphones, transparent mode, bone conduction glasses & Seeing AI, BeMyEyes, Google Lookout, Lazarillo - GPS for traveling, BeMyAI - \textit{``I use outdoors when I am alone, when I am not familiar''} \\ 
\rowcolor{white}
8 & 70 & F & Glaucoma & <10\% & 50 & Wired earphones & BeMyEyes, BeMyAI, \textit{``I don’t use it outdoors, I ask my husband to describe.''} \\ 
\rowcolor{lightgray!40}
9 & 59 & M & Glaucoma, \textit{``I can see partially''} & not known & 47 & Wired earphones, mostly outside, listen to music and radio & BeMyEyes, Lazarillo \\ 
\rowcolor{white}
10 & 67 & M & Muscular degeneration - \textit{``Left eye completely blind, right eye zero detail in vision, very cloudy''} & not known & 35 & Monster open ear, at home for listening to music & Envision AI, Google Lookout, BeMyEyes, BeMyAI \\ 
\rowcolor{lightgray!40}
11 & 31 & F & Macular Degeneration & \textit{``right eye - <2\%, left eye - <10\%''} & 14 & Bluetooth earphones & Lazarillo - \textit{``meter distance is hard to gauge on gps apps''}; TapTapSee - \textit{``For single objects, this app is good, but for a busy picture, it is not''} \\ 
\hline
\end{tabular}%
}
\end{table*}

\subsection{Listening Test Design}
\subsubsection{Scene images}
\label{sec:sceneimages}
We selected eight Vista space images, evenly split between nature and urban environments. The nature scenes included a \textit{sea beach}, \textit{mountains}, \textit{countryside}, and a \textit{reservoir}, while the urban scenes comprised an outside \textit{train station}, \textit{park}, \textit{food court}, and a \textit{street}. Nature scenes are often associated with leisure and relaxation, while urban scenes are relevant to navigation and daily activities \cite{gupta2024imwut}. This balance allowed us to test our framework's adaptability across different types of environment.

\newtext{The human auditory system has a limited capacity to segregate and process multiple sound sources simultaneously \cite{bregman1994auditory}. Scenes with fewer sound sources are generally easier for listeners to parse and understand, while scenes with more sources introduce complexity that requires more cognitive effort to process. To assess our framework's effectiveness in handling different scene complexities, the scene images were also carefully selected on the basis of the density of sonic objects. The sea beach, mountains, train station, and park images represented low-density scenes ($\sim2$ sonic objects), while the countryside, reservoir, food court, and street images represented high-density scenes ($\sim3-4$ sonic objects). The scene images and the details are provided in Table \ref{table:scenes_study1}.
}

\newtext{This variety in scene selection and sonic object density provided a robust test bed for evaluating how well our Scene2Audio framework adapts to different types of auditory environment.}

\begin{table}[h!]
\centering
\centering
\caption{\newtext{Details of the scene images used for the first two studies. Check out the images and generated sounds on our webpage: \url{https://scenetoaudio.github.io/scenetoaudio/\#/section-4}.}}.
\label{table:scenes_study1}
\vspace{-0.3cm}
\resizebox{\columnwidth}{!}{
\begin{tabular}{|c|c|c|c|c|}
\hline
  \textbf{Scene}      & \textbf{Category} & 
  \textbf{Density} & \textbf{Num.~sonic objects} & \textbf{Example generated}\\&&&&\textbf{Noun+Verb Phrase} \\ \hline
  seabeach&nature&low&2&waves crashing,\\
  &&&&wind blowing\\\hline
  mountains&nature&low&2&river flowing\\&&&& wind blowing\\\hline
  countryside&nature&high&4&cows mooing,\\
  &&&&leaves rustling\\
  &&&&church bell ringing\\
  &&&&birds chirping\\\hline
  reservoir&nature&high&4&water flowing\\
  &&&&trees rustling\\
  &&&&wind blowing\\
  &&&&birds chirping\\\hline
  park&urban&low&2&water rippling\\
  &&&&birds chirping\\\hline
  train station&urban&low&2&turnstile beeping\\
&&&&people walking\\\hline
foodcourt&urban&high&3&cutlery clinking\\
&&&&people talking\\
&&&&leaves rustling\\\hline
street&urban&high&3&people walking\\
&&&&vehicle engines idling\\
&&&&traffic signal beeping\\\hline
\end{tabular}}
\end{table}

\subsubsection{Audio Types}
We generated the four audio types - Audio-only, Speech-only\footnote{Eleven Labs text-to-speech engine: {\url{https://elevenlabs.io/text-to-speech}}, was used to generate the speech.}, Overlay, and Overlay Concat, for all the eight scenes (section \ref{sec:sceneimages})\footnote{Listen to the audio samples on our anonymized webpage: \url{https://scenetoaudio.github.io/scenetoaudio/\#/section-4}}.  
To ensure a fair comparison between audio types, we used different scenes for each audio type and counterbalanced the scene-audio type combination between participants. 
Each participant listened to eight audio clips corresponding to the eight different scene images, two of each audio type. We used a within-subject design to compare the four audio types for different Vista spaces. The order of audio types was counterbalanced to avoid sequence bias.

\subsubsection{Study Instructions}
The listening test was conducted in a controlled lab environment. Participants were asked to imagine themselves on a hilltop or rooftop, using a system that converts distant scenes into audio. \newtext{All participants received the same instructions to ensure consistency across conditions.}



\subsection{Measures}
\label{sec:measures}
\noindent\textbf{Overall Preference:} At the end of listening to the eight audio clips, we asked participants to rank their preference regarding the 4 audio types they had heard. We also asked  open-ended questions about their preference. See Appendix (2) for the full details of the questionnaire. 

\noindent\textbf{Additional Criteria:}
Other than overall preference, we asked the participants a set of questions after listening to each audio sample. This questionnaire consisted of evaluating four key criteria: \textit{Comprehension}, \textit{Engagement}, \textit{Immersion}, and \textit{Cognitive Load}. These criteria were chosen based on the aspects of Vista space experience indicated as important in previous works \cite{gupta2024imwut,gonzalez2024investigating,bregman1994auditory}. We carefully adopted questions from existing questionnaires that assess these criteria in the literature. 
\begin{itemize}
    \item \textbf{Comprehension:} We assessed how well the participants understood the sonified scenes through both an open-ended response on scene description based on the audio sample and a self-reported comprehension rating on a 7-point Likert scale. The participants' descriptions were analyzed using an Intent Similarity score, which calculates the cosine similarity between the word embedding in their description and the ChatGPT-generated reference description, checked by the authors (the text used in the Speech-only condition)\footnote{The word embeddings were extracted using python's NLP library spacy's English language model en\_core\_web\_sm}. The score was scaled to the range of 1 to 7 for comparison with other questions.

\item \textbf{Engagement:} We evaluated engagement in the auditory experience on the dimensions of enjoyment, curiosity, imagination, and memorability, using a 7-point Likert scale. 
We use the dimension of enjoyment because according to the Flow Theory \cite{czikszentmihalyi1990flow}, 
when users find an experience enjoyable, they are more likely to be fully absorbed in it, leading to deeper engagement with the content. By evaluating curiosity, we assess whether auditory feedback stimulates the desire of participants to explore and understand the scene further \cite{berlyne1954theory}, indicating a higher quality of engagement. We evaluate imagination as sounds have the potential to stimulate imagination and create vivid mental representations of the scene \cite{steinhaeusser2021comparing, rodero2012see}, which enhances engagement. And, memorability of experiences is related to engagement of emotions and imagination \cite{rodero2019spark, potter2006made}. 

\item \textbf{Immersion:} We measured how deeply participants feel absorbed in the auditory experience based on the Igroup Presence Questionnaire (IPQ) \cite{melo2023IPQ}, which is a standard self-assessment questionnaire to assess how immersed or ``present'' a person feels within a virtual environment. We use a 7-point Likert scale for the four dimensions of immersion - \textit{Presence}, \textit{Spatial Presence}, \textit{Involvement}, and \textit{Experienced Realism}. We assess immersion of the audio content without spatializing sounds. 

\item \textbf{Cognitive Load:} We used NASA-TLX \cite{hart1988development} to measure the cognitive effort required to process the auditory scene.
\end{itemize}

\subsection{Results}
\subsubsection{Overall Preference}
At the end of listening to the four audio types of all the scenes, each participant was asked to rank order the four audio types according to their preference, rank 1 being the most preferred audio type. We assigned a preference score 4.0 to a rank 1 and score 1.0 to a rank 4. The average preference scores are shown in Figure \ref{fig:rankDist}. Overlay is the highest preferred audio type, with a positive inter-rater agreement (Kendall coefficient $\tau = 0.3, p = 0.0$).

\begin{table}[h!]

\begin{minipage}{0.48\textwidth}
\centering
\includegraphics[width=\textwidth]{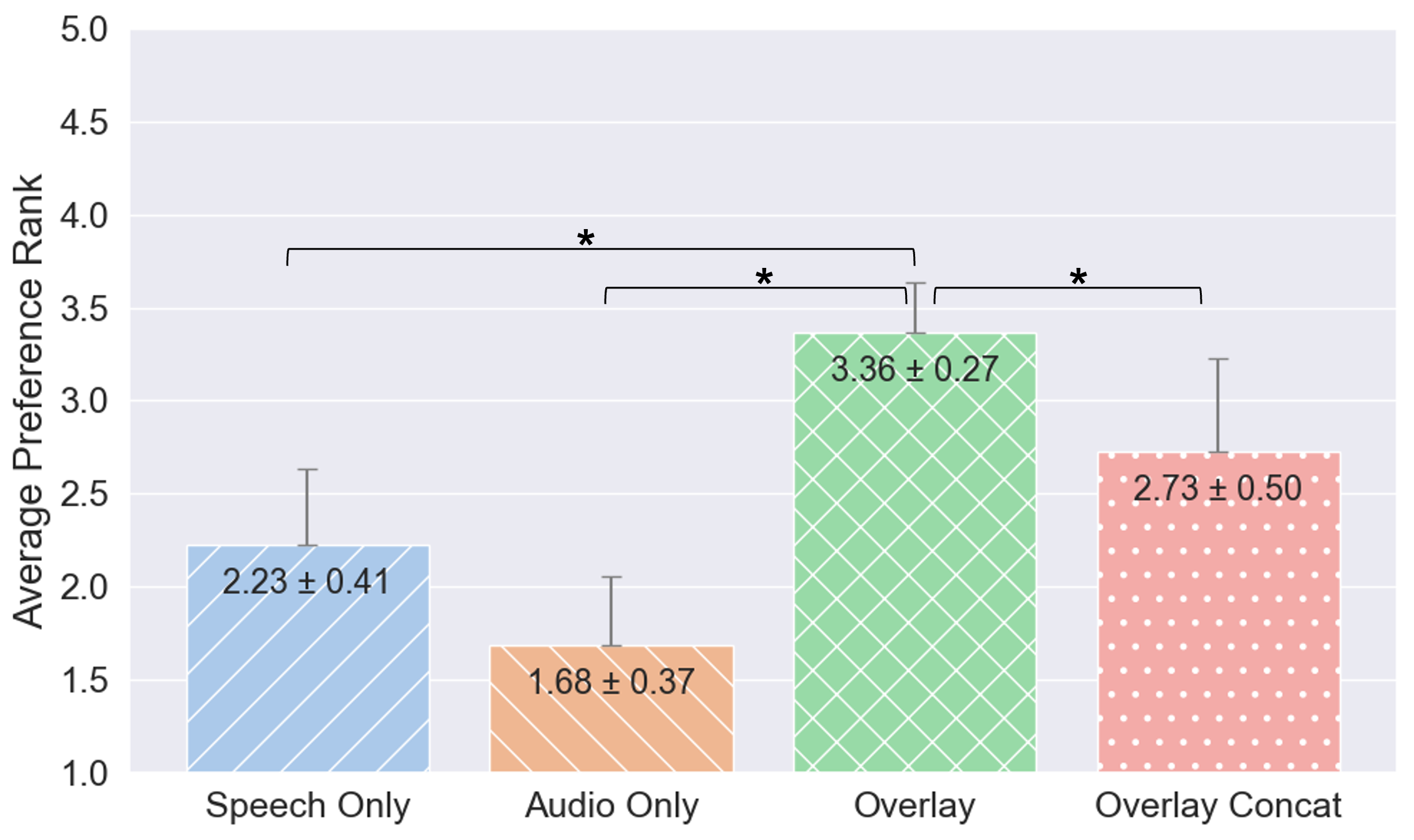}
\vspace{-0.5cm}
\captionof{figure}{Average Rank-order preference score ($\pm$Std Dev) of Audio types. Error bars show 95\% confidence intervals. Pairwise statistical significance using Wilcoxon signed rank test is shown with a \textbf{*} for significantly different pairs ($p<0.05$).}
\Description{
This figure is a bar chart comparing the 'Preference Score' across four different conditions: Speech Only (light blue with diagonal lines), Audio Only (orange with diagonal lines), Overlay (green with crosshatch pattern), and Overlay Concat (red with polka dot pattern). The y-axis represents the preference score, ranging from 1 to 5. Error bars are included to depict 95\% confidence intervals, and asterisks mark statistically significant differences between conditions.

- Speech Only has a mean score of 2.23 ± 0.97.
- Audio Only has the lowest mean score of 1.68 ± 0.89.
- Overlay has the highest mean score of 3.36 ± 0.66.
- Overlay Concat has a mean score of 2.73 ± 1.20.

Significant differences are indicated between:
- Overlay and Audio Only,
- Overlay and Speech Only,
- Overlay and Overlay Concat.
}
\label{fig:rankDist}
\end{minipage}
\hfill
\begin{minipage}{0.48\textwidth}
    \centering
\includegraphics[width=\textwidth]{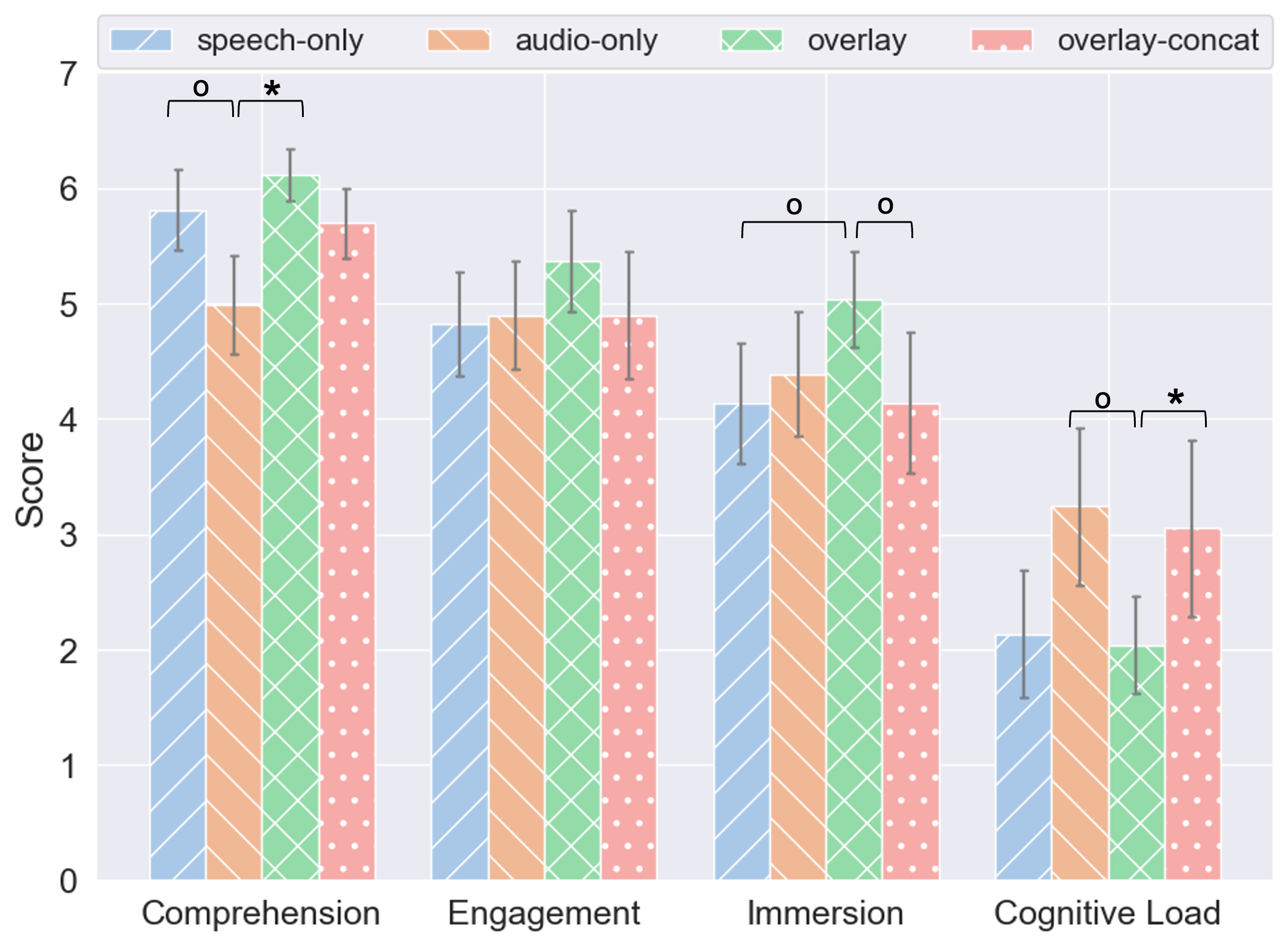}
    \vspace{-0.5cm}
    \captionof{figure}{Overall ratings averaged across questions for each criteria: Comprehension (higher is better),  Engagement (higher is better),  Immersion (higher is better), and Cognitive Load (lower is better). Error bars show 95\% confidence intervals. Post-hoc pairwise statistical significance is shown with a \textbf{*} for significantly different pairs ($p<0.05$), and \textbf{o} for a weakly significant difference between a pair ($p<0.10)$.}
    \Description{
This figure contains bar charts comparing scores of four different conditions that appear as points on x-axis: speech-only (light blue with diagonal lines), audio-only (orange with diagonal lines), overlay (green with crosshatch pattern), and overlay-concat (red with polka dot pattern). The score values are represented on the y-axis, ranging from 1 to 7. Error bars depict 95\% confidence intervals. Asterisks indicate statistically significant differences between conditions.

- The first point on x-axis represents Comprehension criteria ratings with a score comparison of four conditions. The mean scores are:
  - Speech-only: 5.81 ± 0.83
  - Audio-only: 4.99 ± 1.02
  - Overlay: 6.11 ± 0.53
  - Overlay-concat: 5.69 ± 0.72
  Significant differences (p<0.05) are indicated between Overlay and Audio-only conditions, and weakly significant differences (p<0.10) are indicated between Speech-only and Audio-only conditions.

- The second point on x-axis represents Engagement criteria ratings with a score comparison of:
  - Speech-only: 4.82 ± 1.08
  - Audio-only: 4.90 ± 1.13
  - Overlay: 5.36 ± 1.04
  - Overlay-concat: 4.90 ± 1.32
  No significant or weakly significant differences are reported for this criteria.

- The third point on x-axis represents Immersion criteria ratings with score comparison of:
  - Speech-only: 4.14 ± 1.25
  - Audio-only: 4.39 ± 1.29
  - Overlay: 5.03 ± 1.00
  - Overlay-concat: 4.14 ± 1.46
  Weakly significant differences (p<0.10) are indicated between Overlay and both Speech-only and Overlay-Concat conditions.

- The fourth point on x-axis represents Cognitive Load criteria ratings, where lower is better, with score comparisons:
  - Speech-only: 2.14 ± 1.31
  - Audio-only: 3.24 ± 1.63
  - Overlay: 2.04 ± 1.01
  - Overlay-concat: 3.05 ± 1.83
  Significant differences are noted between Overlay and Overlay-Concat, and weakly significant difference is noted between Overlay and Audio-only conditions.

Overall, across all charts, the 'Overlay' condition tends to score the highest in multiple categories, and the statistical significance is marked by asterisks, and weak statistical significance is marked by the symbol o.
}
\label{fig:overallratings}
\end{minipage}
\end{table}

\subsubsection{Qualitative Feedback}
We conducted reflexive thematic analysis \cite{braun2021one} on open-ended responses (Section~\ref{sec:measures}), iteratively refining codes (e.g., merging "Stirs emotions" and "Memorable") to derive themes. Analysis involved familiarization, initial coding,  theme development, review, definition, and reporting. Two researchers resolved discrepancies through discussion, ensuring intercoder reliability.

Three key themes emerged from responses to \textit{``Why did you prefer [highest-ranked sound type]?''}:

\noindent{\textbf{1. Enhanced Immersion through Overlay and Overlay-concat sounds:}} Overlay audio (speech + non-verbal sounds) was praised for \textit{``setting the stage''} (P1) by combining scene-setting sounds with detail-oriented speech. Participants who gave a high rank to overlay and overlay-concat sounds reported increased spatial presence (e.g., \textit{``The sound effects create movement in the scene, and stirs emotion, and imagination''} (P6)).

\noindent\textbf{2. Trade-offs in Directionality vs. Coherence:} While Overlay-Concat provided clear directional cues (\textit{``helps follow the scene step-by-step''} (P3)), its sequential delivery increased cognitive load (\textit{``hard to stitch parts together''} (P5)). Participants advocated for smoother transitions to balance structure and flow.

\noindent\textbf{3. Functional Roles of Speech vs. Non-Verbal Audio:} Responses to \textit{``Roles of speech and audio effects?''} revealed a complementary relationship. 
Non-verbal audio complemented the speech by helping in faster understanding (\textit{``I can comprehend overlay sounds more because I can capture fast with the audio effects''} (P8)), and in making the experience more realistic and easy to imagine (\textit{``Adding sound effects to the description makes the scene more natural, realistic, and easy to visualize''} (P10); \textit{``the distinctive sounds [non-verbal sounds] really helps in imagining the scene and brings you into the scene''}). But sound effects alone, without supporting speech, were ``confusing'' (P1) and ``hard to understand'' (P5). 

Speech, on the other hand, provided \textit{``details [that] guide you''} (P7), but lacked immersion alone, without the sound effects (\textit{``feels stationary''} (P6)).

\subsubsection{Additional Criteria}

The four criteria: Comprehension, Engagement, Immersion, and Cognitive Load, were evaluated across the four different audio types: Speech-only, Audio-only, Overlay, and Overlay Concat. 
We first conducted the Shapiro-Wilk normality test. When the normality assumption was not met for all variables, the results were analyzed using the non-parametric Friedman test for main effect, and the non-parametric Wilcoxon signed-rank test with Bonferroni correction was used for post-hoc analysis. When normality assumption was met for all variables, ANOVA test was used for main effect, and post-hoc pairwise comparisons were conducted using Tukey’s Honestly Significant Difference (HSD) test, which adjusts for multiple comparisons. The overall scores, i.e.~average scores across questions for each criterion is visualized in Figure \ref{fig:confidence_pleasantness}, with statistically significant comparisons (Wilcoxon signed rank test, $p<0.05$) marked with $*$. The individual questions for each criteria are also analyzed for statistical significance in the following sub-sections. The detailed average values for each question (mean $\pm$ standard deviation) are provided in Table \ref{tab:audio_scores} in Appendix (3). 
\begin{itemize}[leftmargin=0pt, itemindent=*]
    \item \textbf{Comprehension: }
Across the overall comprehension scores for the four audio types (at least one group was not normally distributed), Friedman test ($\chi^2(3) = 8.2, p = 0.04$) revealed a statistically significant main effect. The post-hoc Wilcoxon signed-rank test with Bonferroni correction showed  Overlay ($6.11$$\pm$$0.53$) significantly outperformed Audio-only ($4.99$$\pm$$1.02$) ($p = 0.006$), and Speech-only ($5.81$$\pm$$0.83$) and Audio-only showed weakly significant difference ($p = 0.06$)(Figure \ref{fig:overallratings}). Figure \ref{fig:comprehension} in Appendix shows the average ratings for each of the questions of the Comprehension criteria.

For the Intent Similarity question under Comprehension criteria, Friedman test did not show a statistically significant main effect  ($\chi^2(3) = 6.9, p = 0.07$). However, for the Self-reported Comprehension Rating question, for which the variables were normally distributed, a one-way ANOVA ($F(3,84) = 7.17$, $p = 0.0002$) showed a statistically significant main effect across audio types. The post-hoc Tukey HSD test revealed that overlay, overlay-concat, and speech-only significantly outperformed audio-only ($p_{adj} = 0.0002$, $p_{adj} = 0.017$, and $p_{adj} = 0.003$ respectively). Overlay achieving the highest mean score of $6.23 \pm 0.87$, but no significant differences were found between overlay, overlay-concat, and speech-only. 


\item\textbf{Engagement: }
For the Overall Engagement score across the four audio types (all groups normally distributed), 
one-way ANOVA 
did not show a statistically significant main effect 
($F(3,84)=1.032$, $p = 0.383$). None of the pairs showed any post-hoc weakly significant difference either.
Figure \ref{fig:engagement} in Appendix shows the average ratings for each of the questions of the Engagement criteria. 

For the Enjoyability Rating question, Friedman Test ($\chi^2(3) = 9.76, p = 0.02$) showed a statistically significant main effect across audio types. The post-hoc Wilcoxon signed-rank test with Bonferroni correction revealed that overlay was significantly more enjoyable than audio-only ($p = 0.01$). The other three questions on Curiosity, Imagination, and Memorability, showed no significant differences in main effect. The Curiosity related question saw the Audio-only condition scoring highest, while for the other three questions, Overlay condition scores the highest.


\item\textbf{Immersion: }
For overall Immersion scores (all groups normally distributed), 
the one-way ANOVA did not show a statistically significant main effect  
($F(3,84) = 2.477$, $p = 0.0669$)
. Post-hoc Tukey HSD showed weakly significant difference between speech-only and overlay ($p = 0.09$), as well as overlay-concat and overlay ($p = 0.09$), showing overlay was somewhat more immersive than speech-only and overlay-concat, aligning with the qualitative responses. 
Figure \ref{fig:immersionquestions} in Appendix shows the average ratings for each of the questions of the Immersion criteria.

The Overlay condition consistently scored the highest across all factors of Immersion criteria - Presence (``sense of being present in the scene''), Spatial Presence (``scene surrounded me''), Involvement (``not aware of real environment''), and Experienced Realism (``seems like the real world'') (please refer to Table \ref{tab:audio_scores} in Appendix). Notably, Experienced Realism question (normally distributed) showed significant main effect in one-way ANOVA ($F(3,84) = 3.28$, $p = 0.02$), with Overlay audio type rated significantly higher than Speech-only ($p_{adj} = 0.03$). 

\item\textbf{Cognitive Load: }
Regarding Cognitive Load, lower scores indicate better performance (at least one group was not normally distributed). 
Friedman test 
indicated statistically significant main effect in Cognitive Load across the audio types ($\chi^2(3) = 9.8, p = 0.02$). Post-hoc Wilcoxon signed-rank test with Bonferroni correction showed overlay had significantly lower cognitive load than overlay-concat ($p = 0.04$), and a weakly significant lower cognitive load than audio-only ($p = 06$).
Figure \ref{fig:cognitiveload} in Appendix shows the average ratings for each of the questions of the Cognitive Load criteria. 

The Overlay audio type again showed the best performance with the lowest scores for all the questions. Mental Demand (normally distributed) had significant main effect through ANOVA ($F(3,84) = 4.62$, $p = 0.005$). Post-hoc with Tukey HSD showed that both overlay and speech-only were significantly less mentally demanding than audio-only ($p_{adj} = 0.038$ and $p_{adj} = 0.04$ respectively) as well as overlay-concat ($p_{adj} = 0.04$ and $p_{adj} = 0.04$ respectively). No differences were found between speech-only and overlay in terms of mental demand.  

The question on Difficulty in Performance (normally distributed) showed significant main effect through ANOVA ($F(3,84) = 3.17$, $p = 0.03$) with post-hoc Tukey HSD showing that Overlay was significantly less difficult than audio-only ($p_{adj} = 0.03$), while the other pairwise comparisons showed no significant differences. 

The question on  Effort (not normal) showed significant main effect through Friedman test (($\chi^2(3) = 11.9, p = 0.008$)). Post-hoc with Wilcoxon signed rank test with Bonferroni correction showed that overlay required significantly less effort than overlay concat ($p = 0.04$), while the other pairwise comparisons showed no significant differences. 

The questions on Physical Demand, Temporal Demand, and Frustration did not show significant main effect. 

\end{itemize}

\section{In-the-Wild Study}
Building on the controlled lab evaluation, we developed a mobile app and a study to assess its real-world effectiveness in enhancing scene awareness and experience for BLV users. This study investigates how participants engage with the app in their daily outdoor environments in their own time, evaluating its usage in real-world contexts, i.e.~ecological validity. Specifically, our aim was to explore the following research questions:
\begin{itemize}
    \item Which of the four sonification strategies is preferred by BLVs in real-world contexts? What are the real-world contexts in which sound effects are desirable by BLVs?
\item What is the effect on the user experience when the app is used over a period of time (over a week)?
\item What are the key points for improvement?
\end{itemize}
\subsection{Mobile App Design}
The design of the Sonic Vista app prioritizes accessibility and ease of use for BLV users, informed by user behavior research. To ensure accessibility, the interface employs high-contrast colors. Functional elements, such as buttons, are positioned at the screen's corners and bottom, aligning with common interaction patterns among blind users. The buttons are large and the overall interaction flow is simplified to enhance convenience.

\begin{figure}
    \centering
    \begin{subfigure}[b]{0.45\columnwidth}
   \centering
   \includegraphics[width=\linewidth]{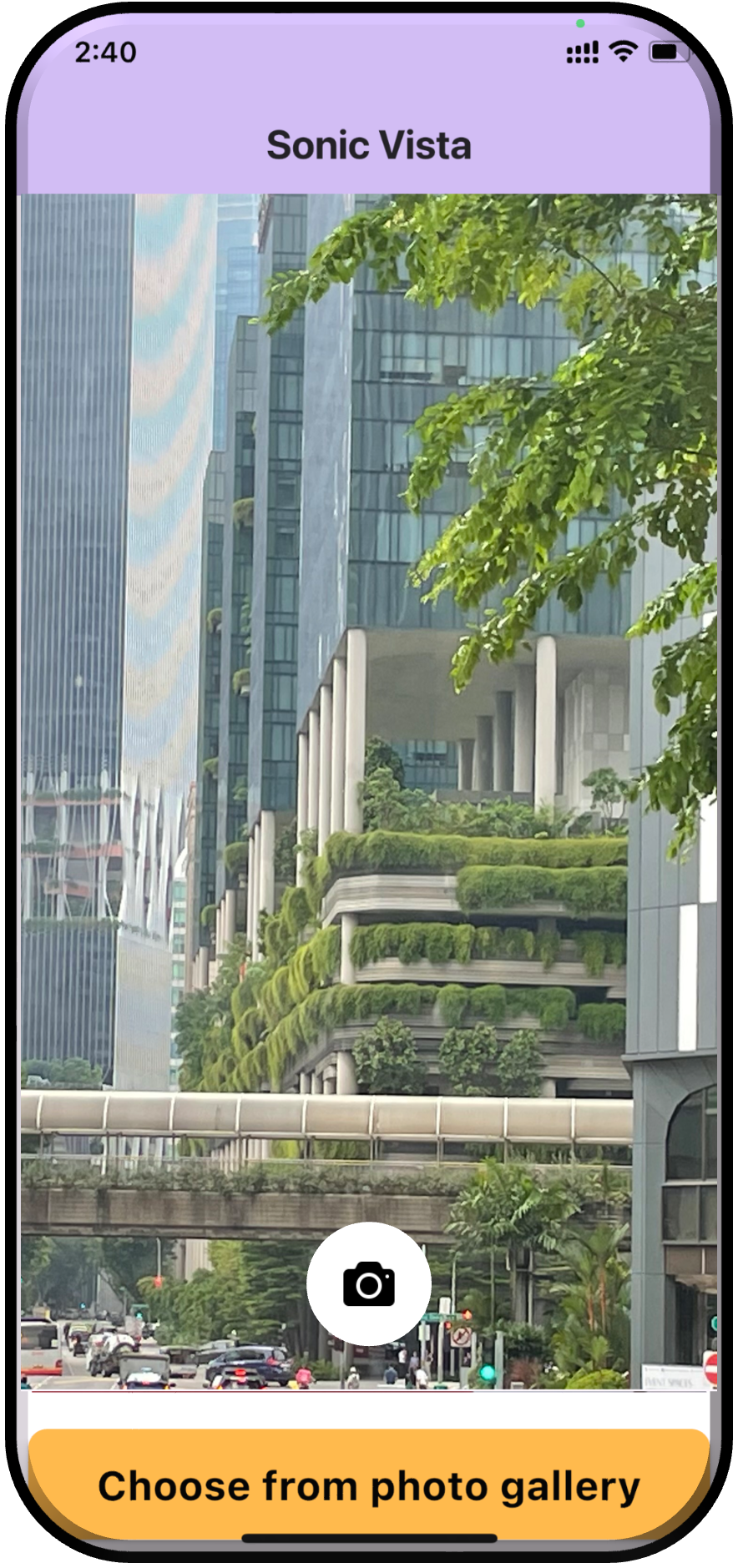}
   \vspace{-0.1cm}
   \caption{}
   \label{fig:f1} 
\end{subfigure}
\begin{subfigure}[b]{0.45\columnwidth}
   \centering
   \includegraphics[width=\linewidth]{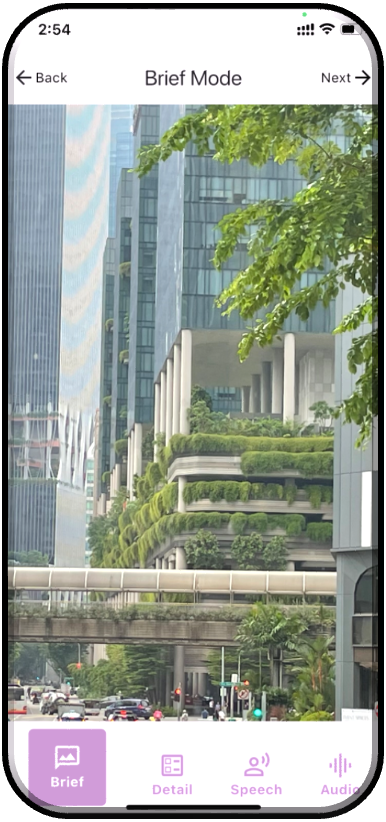}
   \vspace{-0.1cm}
   \caption{}
   \label{fig:f1} 
\end{subfigure}
\vspace{-0.3cm}
    \caption{Mobile App Screenshots (a) Take/Select a photo, (b) Audio playback page}
    \Description{The figure shows two screenshots of a mobile application.
(a) The first screenshot is labeled “Sonic Vista.” It shows a photo of tall buildings with greenery on multiple levels, along with a road and traffic at the bottom. At the bottom of the screen, there is a large yellow button labeled “Choose from photo gallery.” In the middle, there is also a circular camera button with a camera icon.
(b) The second screenshot is labeled “Brief Mode.” It shows the same photo of the buildings with greenery. At the top, navigation arrows and the text “Brief Mode” are visible. At the bottom, there are four menu options in a purple bar: Brief, Detail, Speech, and Audio, each with an icon above the text. The Brief option is highlighted.}
    \label{fig:app}
    \vspace{-0.5cm}
\end{figure}

The interface features a large camera preview for easy framing, with a high-contrast "Choose from Gallery" button at the bottom (Figure \ref{fig:app}(a)). Users can tap anywhere on the camera preview or use volume buttons for photo capture, with pinch-to-zoom available for framing adjustments. Photos are uploaded to the Scene2Audio backend via a Flask API on a cloud server, which processes the image and returns four audio files back to the app, typically within 15 seconds. 
On the audio playback screen, the current mode is shown at the top along with the 'Back' and 'Next' buttons for navigation. The captured image appears centrally to provide visual context, and the bottom navigation bar hosts the four audio files in the modes labeled Brief, Detail, Speech, and Audio, with the active mode highlighted. These modes correspond to the four strategies evaluated in our previous lab study: Overlay (Brief), Overlay Concat (Detail), Speech-only (Speech), and Audio-only (Audio). According to the lab study  indicating that it is the preferred one, the Brief mode plays first. Then the user can switch to the other modes with either taps or swipes, enabling easier navigation. The user can hit the Next button once all four modes have been played at least once. A built-in questionnaire (discussed in Section \ref{sec:questionnaires}) follows, collecting user ratings, preferences, and feedback, with the images and the user feedback data stored in Google Cloud for analysis.

The app is designed with VoiceOver support, incorporating custom gestures, clear button labels, and voice guidance to ensure seamless navigation. For instance, on the screen capture , a user tapping the camera preview area hears a prompt saying, “Take a picture, double-click to take a photo.” Audio playback screen controls also adopt double-tap functionality, ensuring consistency with VoiceOver interactions in other apps such as BeMyEyes and SeeingAI. \newtext{The in-app questionnaires were enabled with the phone's in-built speech-to-text for better accessibility.}

The app was developed on Flutter and was made available for both Android and iOS phones.

The backend consisting of the four audio generation pipelines was deployed on a Linux server with a GPU ( NVIDIA GeForce RTX 3090 GPU with 24GB of VRAM) through Flask API. 

\subsubsection{Latency Analysis}
\newtext{To assess latency variability and rate limits of the app, we benchmarked our Scene2Audio pipeline by running the system 50 times sequentially. Each run generated the four audio outputs corresponding to the four modes from an input image. The average end-to-end latency per image was 15.9 s (SD = 0.93 s), with a range of 14.1-17.5 s, indicating low variance across repeated generations.}

\subsection{Study Design}
\subsubsection{Participants}
For the mobile app study, we first conducted a pilot study with P1, which helped us to ensure that the buttons and features of the app, especially those related to accessibility, were functional, so the app was easy to use. We then recruited 7 out of the 11 BLV participants from the controlled lab study (Table \ref{tab:pvi_details}), based on their availability: P2, P3, P4, P5, P7, P9, and P11. Three are iOS users, and four are Android users. We obtained ethics approval from the Institute Review Board (IRB) for this study before conducting it.

\subsubsection{Method}
Participants were asked to install the app on their smartphones and use it independently over a week. 
Participants were asked to use the app outdoors, when they were seated and at leisure. The task was to take pictures through this app, listen to all the four audio modes, and fill in the questionnaire on the "Next" page for every image they clicked. \newtext{We adopted a full-exposure design so that every participant could experience and compare all four modes on the scenes they encountered in their own time, enabling consistent, scene-matched feedback across modes. While this structure provides controlled comparability, it also means that participants did not switch modes freely as they might in everyday use. Future iterations can explore more naturalistic, self-directed mode switching once the baseline preferences are established.}

The participants were asked to submit at least 10 images along with their questionnaire over at least one week period. 
They were also requested to use their headphones when they listened to the sounds to have a better experience.
After they submitted 10 images, a telephonic semi-structured interview was conducted to understand their overall impressions and reflections on the experience.

\subsubsection{Questionnaires}
\label{sec:questionnaires}
Three different in-app questionnaires were used to capture user preferences, experiences, and reflections. These three questionnaires are provided in the Appendix.

\textit{Mode Preference Questionnaire: }This in-app questionnaire appeared after each photo submission. It asks participants which of the four audio modes they prefer for the given scene and why. It also captures which modes they find most clear and enjoyable, and which is least effective.

\textit{User Experience Questionnaire (UEQ): }The UEQ is a standardized questionnaire to measure how participants' experience evolves over time. It was presented once every three photo submissions after the Mode Preference questionnaire. Since UEQ is designed to track novelty effects in new technologies, it helps to assess whether user ratings change when the novelty effect of the app decreases.

\textit{Post-Study Interview: }After the study, i.e.~after the task of 10 photo submissions via the app was completed , the participants completed a telephonic semi-structured interview. This final questionnaire builds on patterns observed in previous responses, encouraging participants to reflect on their overall experience, share qualitative feedback, and discuss possible improvements.

\subsection{Results}
\newtext{Each of the seven participants clicked and submitted at least 10 images during the study, for a total of 77 images. The photos covered a mix of natural landscapes, urban environments, indoor settings, and close-up objects. Across all participants, 58 images were labelled as outdoor (e.g., streets, parks, waterfronts, building facades) and 19 as indoor (e.g., inside rooms, and corridors).}
\newtext{Most of the outdoor scenes were clicked from a vantage point covering a landscape, therefore can be considered as vista scenes. There were 6 (P2), 4 (P3), 9 (P4), 8 (P5), 12 (P7), 8 (P9), and 11 (P11) outdoor photos.} The rest were indoor scenes that included one or a few objects in focus and in close range. Some images clicked by the participants and their feedback after listening to the four audio types on the app corresponding to the image are provided in Table \ref{tab:inthewildresponses}, and the complete extracted data is provided here\footnote{\url{https://rb.gy/latgtg}}. Depending on their availability, the participants took variable number of days to finish the task - 30 days (P2), 11 days (P3), 7 days (P4), 7 days (P5), 12 days (P7), 9 days (P9), and 15 days (P11). The photos were taken spread out over the whole period of time, as can be observed in the extracted data$^{16}$. 
\begin{table*}[]
\caption{Subset of in-the-wild user study participant responses. To refer to all the responses, see $^{16}$}
\label{tab:inthewildresponses}
\vspace{-0.3cm}
\centering
\resizebox{0.9\textwidth}{!}{%
\begin{tabular}{@{}m{0.5cm}m{3cm}m{1cm}m{1cm}m{1cm}m{1cm}m{1cm}m{3cm}m{3cm}m{1cm}@{}}
\toprule
PID & Image & Clearest info & Least clear info & Most enjoyable & Least enjoyable & Most preferred & Why & \begin{tabular}[c]{@{}c@{}}What did you\\ want to know\end{tabular} & \begin{tabular}[c]{@{}c@{}}Did you get\\ the info\\ you wanted\end{tabular} \\
\midrule
P4 & \includegraphics{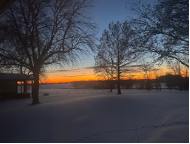} & Detail & Audio & Detail & Audio & detail & gave the most detail and had some sound effects enhance the experience. & description of the scene & True \\
P2 & \includegraphics{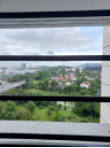} & Detail & Audio & Detail & Audio & detail & because it gives me clearest possible picture in my mind &  & True \\
P3 & \includegraphics{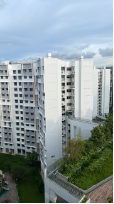} & Detail & Brief & Detail & Brief & detail & Details and precise location that I can visualise & I wish it could describe colours of the environment as well & True \\
P5 & \includegraphics{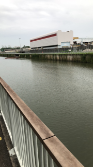} & Detail & Audio & Detail & Speech & detail & Canal at the park. & I especially enjoyed the sounds of the birds, water, etc. However, the app failed to inform me if there were fishes in the water. & True \\
P7 & \includegraphics{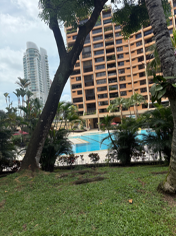} & Detail & Audio & Detail & Speech & detail & I got all the details, especially the pool & the swimming pool & True \\
P11 & \includegraphics{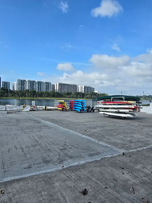} & Detail & Audio & Brief & Audio & Brief & It was concise and easy to understand. It gave a good overview of the scene ahead of me. &  & False \\
P9 & \includegraphics{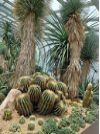} & Detail & Audio & Detail & Audio & detail & more description &  & True \\
\bottomrule
\end{tabular}}
\end{table*}
\subsubsection{Mode Preference}
\begin{figure*}
    \centering
    \includegraphics[width=\linewidth]{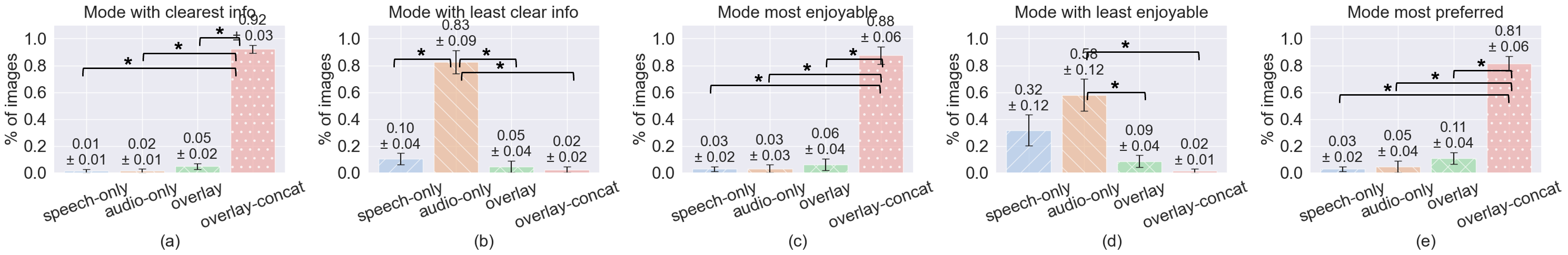}
    \caption{Mode (in percentage) that : (a) provided the clearest info; (b) provided the least clear info; (c) provided the most enjoyable experience; (d) provided the least enjoyable experience; and (e) was overall preferred the most.}
    \Description{This figure, titled Preference Analysis, is composed of five bar charts that show user preferences for different modes of interaction. Each chart compares four modes: speech-only, audio-only, overlay, and overlay-concat. The y-axis on each chart represents the percentage of responses for each mode. Error bars indicate a 95\% confidence interval, and asterisks above the brackets show statistically significant differences (p<0.05).
    Key Findings by Chart
Chart (a): Mode with clearest info. The overlay-concat mode was overwhelmingly rated as providing the clearest information, with a score of 92\%, which is significantly higher than the other modes. The other three modes (speech-only, audio-only, and overlay) each scored around 1-5\%, with no significant difference among them.
Chart (b): Mode with least clear info. In contrast, the audio-only mode was rated as providing the least clear information by 83\% of users, significantly higher than all other modes. The overlay-concat and overlay modes were rated as least clear by only 5\% of users, while the speech-only mode was rated as least clear by 10\%.
Chart (c): Mode most enjoyable. The overlay-concat mode was also rated as the most enjoyable experience, receiving a score of 88\%. This is significantly higher than the other three modes, which all scored below 6\%. There were no significant differences in ratings among the speech-only, audio-only, and overlay modes for this category.
Chart (d): Mode with least enjoyable experience. The audio-only mode was rated as the least enjoyable by 58\% of users, significantly higher than the other three modes. The speech-only mode was rated as least enjoyable by 32\% of users, while the overlay and overlay-concat modes were rated as least enjoyable by 9\% and 2\%, respectively.
Chart (e): Mode most preferred overall. The overlay-concat mode was the most preferred mode overall, with a preference score of 81\%. This is significantly higher than all other modes. The speech-only, audio-only, and overlay modes were preferred by 3\%, 5\%, and 11\% of users, respectively, with no significant difference in preference among them.
Across all five charts, the overlay-concat mode consistently received the most positive ratings, excelling in clarity, enjoyability, and overall preference. Conversely, the audio-only mode was consistently rated the worst, providing the least clear and least enjoyable experience.
    }
    \label{fig:modequestionnaire}
\end{figure*}

From the mode preference questionnaire, we calculated the percentage of times each mode was rated as providing ``clearest info'', ``least clear info'', ``most enjoyable'', ``least enjoyable'', and ``most preferred'', as shown in Figure \ref{fig:modequestionnaire}. Using the Shapiro-Wilk test, we found that normality assumption was not met for all five questions. So we conducted a nonparametric Friedman test for main effect, and a Wilcoxon signed rank test with Bonferroni correction for post hoc analysis. 

Regarding the mode that provides the clearest info, the Friedman test shows significant differences ($\chi^2(3) = 17, p = 0.0007$), with post-hoc tests showing overlay-concat significantly better than all the other modes ($p<0.02$). Friedman results ($\chi^2(3) = 15.9, p = 0.001$) and post-hoc indicate that audio-only is significantly less preferred in terms of clarity compared to speech-only, overlay, and overlay-concat ($p<0.03$). In terms of enjoyability, Friedman results ($\chi^2(3) = 17.3, p = 0.0006$) and post-hoc show that overlay-concat significantly outperforms all the other modes ($p<0.02$). The Friedman results ($\chi^2(3) = 14.06, p = 0.003$) and post-hoc show that audio-only is significantly less enjoyable compared to overlay and overlay-concat ($p<0.05$). Speech-only is also found to be significantly less enjoyable than overlay-concat ($p<0.03$). Finally, the overall  preferred mode is  overlay-concat according to the Friedman  ($\chi^2(3) = 16.65, p = 0.0008$) and post-hoc ($p<0.02$) results. 

To sum up, most users engaged in outdoor spaces with complex scenes including multiple objects, as can be observed in Table \ref{tab:inthewildresponses}. The participants significantly preferred the overlay-concat mode (i.e.~Detail mode), in terms of clarity of information, enjoyment of the experience, and overall preference. In simpler outdoor scenes with fewer objects in focus, such as a road, an animal, or a flower, users tended to prefer brief mode, i.e.~overlay, as they only needed a concise description. 

\subsubsection{Indoor versus Outdoor Scenes}
\begin{figure*}
    \centering

    \begin{subfigure}[b]{0.32\textwidth}
        \centering
        \includegraphics[width=\linewidth]{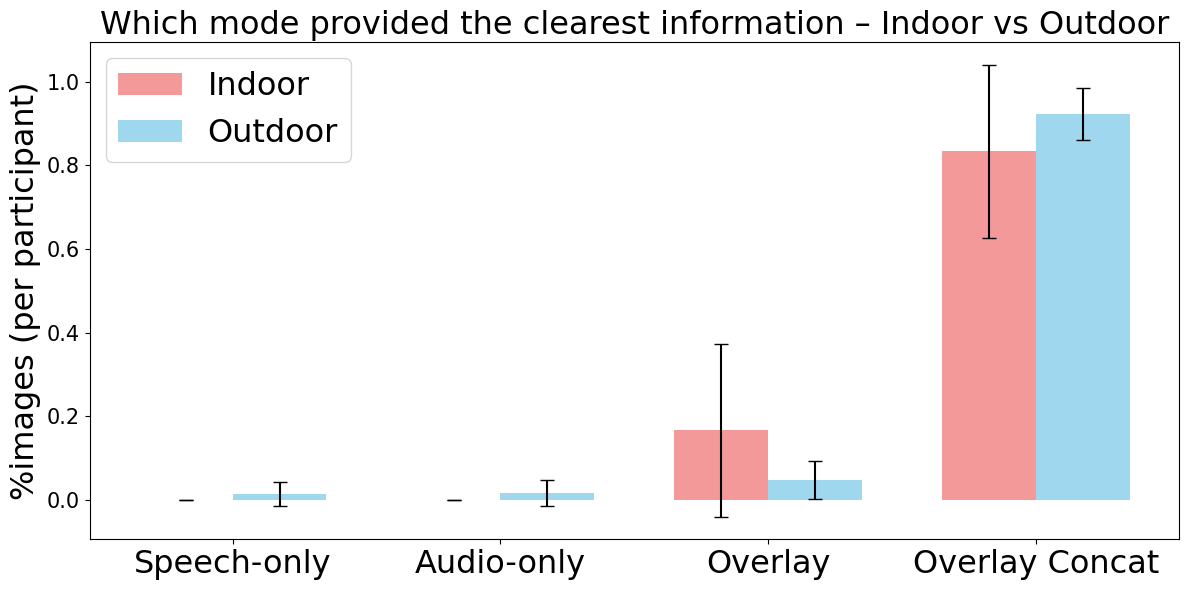}
        \caption{}
        \label{fig:clearest_indoor_outdoor}
    \end{subfigure}
    \hfill
    \begin{subfigure}[b]{0.32\textwidth}
        \centering
        \includegraphics[width=\linewidth]{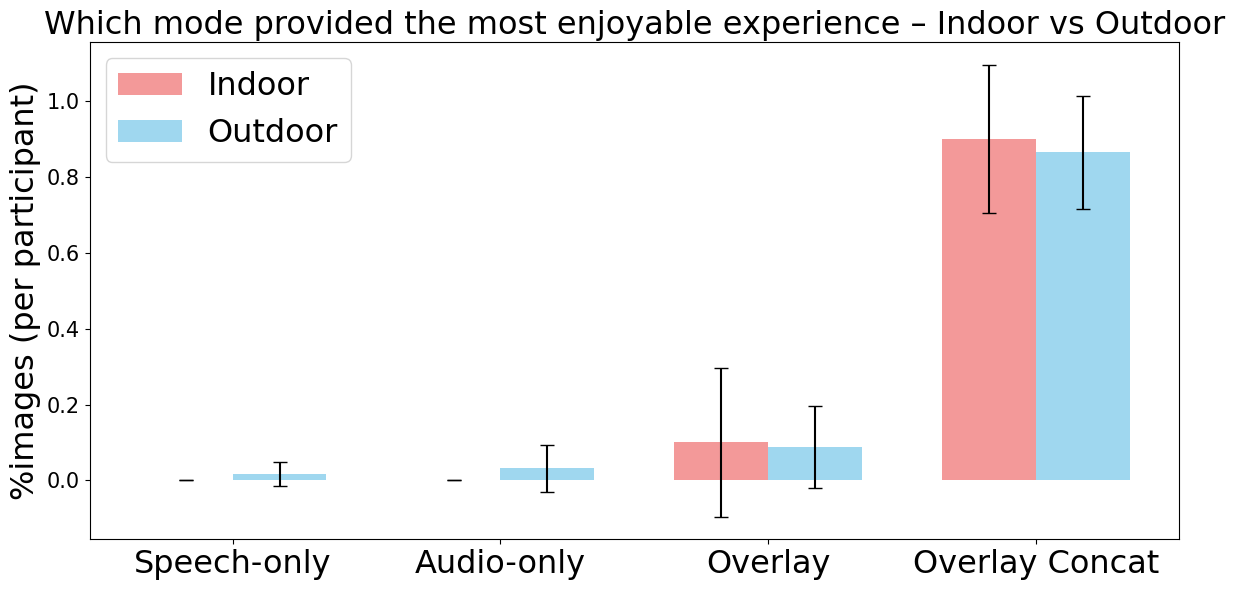}
        \caption{}
        \label{fig:enjoyable_indoor_outdoor}
    \end{subfigure}
    \hfill
    \begin{subfigure}[b]{0.32\textwidth}
        \centering
        \includegraphics[width=\linewidth]{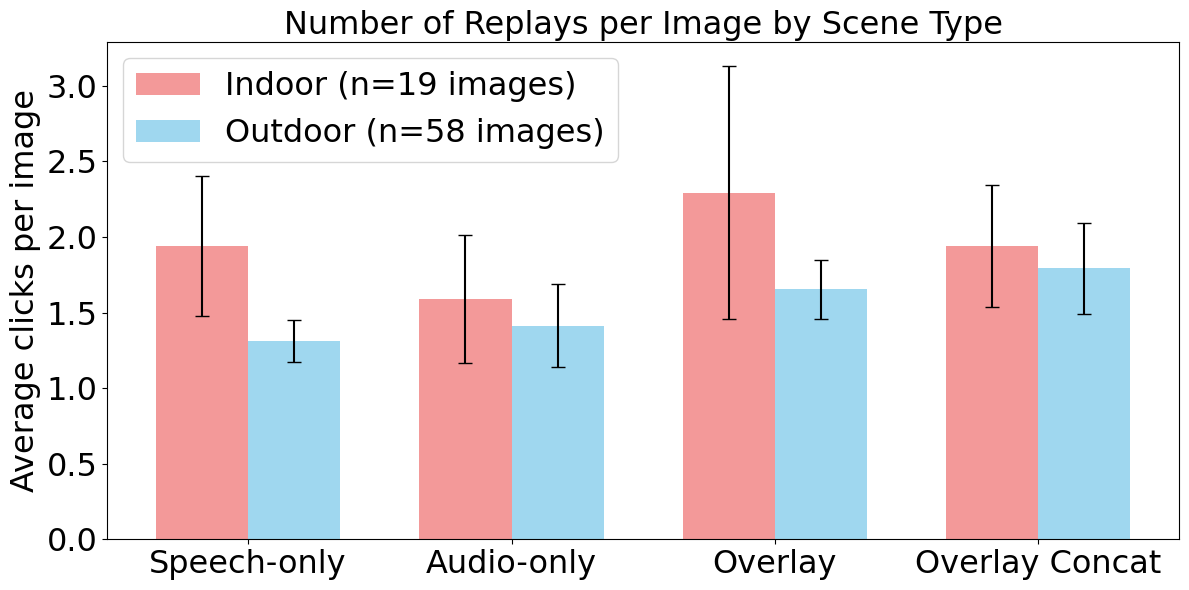}
        \caption{}
        \label{fig:clicks_per_mode}
    \end{subfigure}
\vspace{-0.3cm}
    \caption{
    Indoor vs.~outdoor scenes comparison of (a) mode providing the clearest information,  
    (b) mode rated as most enjoyable, and  
    (c) average clicks per image for each mode.  
    }
    \Description{
Figure 7 compares participant responses for indoor versus outdoor vista scenes across four audio modes: Speech-only, Audio-only, Overlay, and Overlay Concat. 
Panel (a) shows which mode provided the clearest information. Overlay Concat is overwhelmingly chosen for both indoor and outdoor scenes, with very low selections for all other modes. 
Panel (b) shows which mode was rated most enjoyable. Again, Overlay Concat dominates for both scene types, with Overlay receiving a small number of votes and the other modes near zero. 
Panel (c) shows the average number of replays per image. Speech-only and Overlay require the most replays, while Audio-only and Overlay Concat require fewer replays, with indoor scenes generally showing slightly higher replay counts than outdoor scenes. 
Overall, the figure illustrates consistent preferences for Overlay Concat in clarity and enjoyment, and moderate differences in replay behavior across scene types.}
   \label{fig:indoor_outdoor_comparison}
\end{figure*}

\newtext{When we separate the analysis by scene type (indoor vs outdoor), overlay-concat remains the dominant choice for “clearest information” and “most enjoyable experience” in both categories (Fig.~\ref{fig:indoor_outdoor_comparison} (a), (b)). Speech-only and audio-only are rarely chosen as clearest or most enjoyable in either setting. Click-log analysis further shows that overlay and overlay-concat are replayed more often per image than speech-only and audio-only for both indoor and outdoor scenes (Fig.~\ref{fig:indoor_outdoor_comparison} (c)), suggesting that participants actively revisit the richer modes beyond the single mandatory exposure.}


\newtext{Analysis of the free-text “Why” responses further explains the preference patterns. Across both indoor and outdoor scenes, the dominant justification for choosing a mode was the desire for more detail and the ability to form a clearer mental picture of the scene (e.g., “\textit{gave the most detail and had sound effects that enhanced the experience},” “\textit{clearest possible picture in my mind},” “\textit{details and precise location that I can visualise}”). Participants frequently emphasised spatial layout and relative positions—such as knowing where trees, buildings, pools, or objects were located, which was often tied to navigation, safety, or practical understanding of the environment. Outdoors, participants also appreciated naturalistic sound cues (e.g., birds, water) for enhancing immersion, whereas brief summaries were preferred only for simple scenes where a high-level overview sufficed. Several comments also noted missing or incorrect details (e.g., absent lane markings, building types, field slopes, signage text, or occasional hallucinations), highlighting expectations for more complete and reliable scene information. Importantly, these qualitative justifications showed no clear pattern with scene crowd level or sound density, consistent with our stratified quantitative analysis, which revealed minimal variance of preference of audio modes across indoor or outdoor scene types.}
\subsubsection{User Experience}
\begin{figure}
    \centering
    \begin{subfigure}[b]{\columnwidth}
   \centering
   \includegraphics[width=\linewidth]{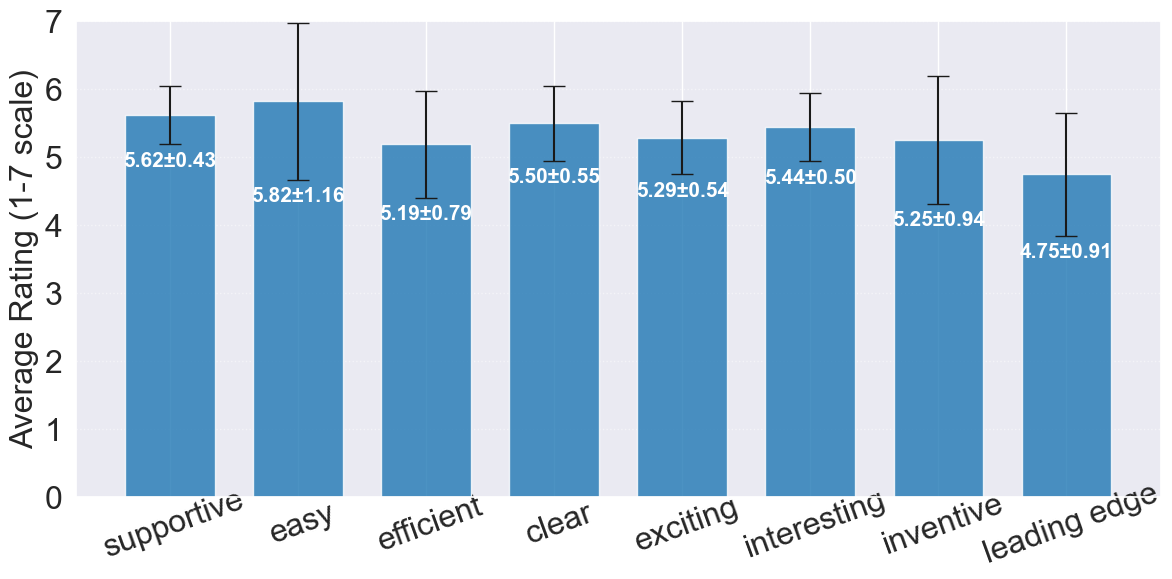}
   \vspace{-0.2cm}
   \caption{}
   \label{fig:f1} 
\end{subfigure}
\begin{subfigure}[b]{\columnwidth}
   \centering
   \includegraphics[width=\linewidth]{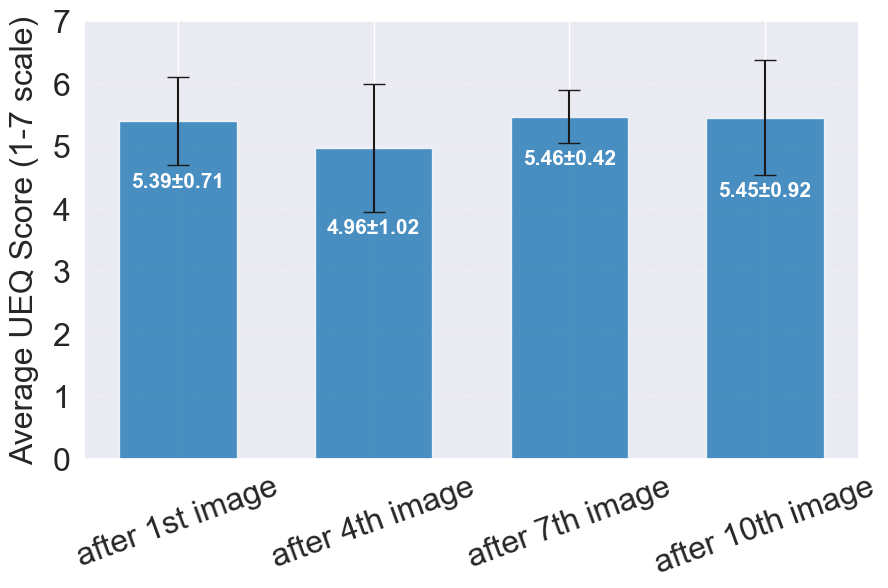}
   \vspace{-0.2cm}
   \caption{}
   \label{fig:f1} 
\end{subfigure}
\vspace{-0.3cm}
    \caption{User experience questionnaire results, (a) Average rating ($\pm$ std.) for each question across all participants, (b) Average UEQ scores ($\pm$ std.) over time across all participants.}
    \Description{This figure is composed of two sub-figures, both displaying results from a User Experience Questionnaire (UEQ). Each chart uses bar graphs to show average scores on a 1 to 7 scale, with error bars representing the standard deviation.
    
    Sub-figure (a): UEQ Results
This chart shows the average rating for eight different questions, or qualities, of the user experience.

Supportive: The highest score, with an average of 5.62±0.43.

Easy: A score of 5.82±1.16, which is the highest average rating in this chart.

Efficient: A score of 5.19±0.79.

Clear: A score of 5.50±0.55.

Exciting: A score of 5.29±0.54.

Interesting: A score of 5.44±0.50.

Inventive: A score of 5.25±0.94.

Leading Edge: The lowest average score, with 4.75±0.91.

Overall, the ratings for all qualities are above 4.5 on the 7-point scale, indicating a generally positive user experience. The qualities rated highest were "Easy" and "Supportive," while the lowest rating was for "Leading Edge."

Sub-figure (b): UEQ Results Over Time
This chart shows how the average UEQ score changes at different points in the user's interaction with the system. The x-axis marks the different points in time: "after 1st image," "after 4th image," "after 7th image," and "after 10th image."

After 1st image: The average score is 5.39±0.71.

After 4th image: The average score is 4.96±1.02. This is the lowest score, representing a slight dip in user experience.

After 7th image: The average score is 5.46±0.42. This score shows an improvement, nearly reaching the initial score.

After 10th image: The average score is 5.45±0.92, which is nearly identical to the score after the 7th image.

The results suggest that the user experience starts strong, dips slightly in the middle of the task, and then recovers and remains stable towards the end. The scores are consistently above 4.5, indicating a positive user experience throughout the process.
    }
    \vspace{-0.5cm}
    \label{fig:ueqresults}
\end{figure}

The User Experience Questionnaire (UEQ) results show distinct patterns across different aspects of the interface and over time. Figure \ref{fig:ueqresults}(a) displays the average ratings for each UEQ dimension, with particularly high ratings for ``easy'' (M = 5.82 ± 1.16) and ``clear'' (M = 5.50 ± 0.55) categories, indicating that participants find the interface intuitive. Slightly lower but still positive ratings were observed for  ``leading edge'' (M = 4.75 ± 0.91) and ``efficient'' (M = 5.19 ± 0.79) dimensions, suggesting room for improvement in response time.

Analysis of user experience score over the duration of the study (Figure \ref{fig:ueqresults} (b)) shows a stable user experience throughout the week with no significant differences between the time intervals. The initial score after the first image (M = 5.39 ± 0.71) remains consistent until the end (M = 5.45 ± 0.92 for the tenth image evaluation). This indicates that even after the first few uses when the novelty of the app usually wears off, the UEQ score still remained consistent. 

\subsubsection{Qualitative Feedback}
Semi-structured post-study interviews with the seven participants show nuanced perspectives on the app's usability, sound design, and real-world utility. Thematic analysis of the interviews reveal four dominant themes, each with implications for future iterations of the technology.

\noindent \textbf{1. Preference for Detailed Descriptions in Real-World Contexts:} Participants consistently favor the detail mode (rich scene descriptions) over brief summaries, which is not consistent with the lab study findings where brevity was preferred due to perceived cognitive load. This shift is attributed to contextual differences: in field use, participants look for concrete information about scenes they partially understand. For instance, P4 explained, \textit{``In the wild, I take pictures to get more details—I already have a rough idea of the scene. Those extra descriptions add to the experience.''} Others, like P9, rejected the brief mode as \textit{``too obvious,''} citing examples where it failed to highlight salient features (e.g., calling a jar simply ``a jar''). P5 noted the lab-field disparity stemmed from differing goals: lab tasks require mental image construction, while real-world use consists more in demanding additional details about familiar environments.

\noindent \textbf{2. Sound Effects: Balancing Immersion and Precision:} 
Sound effects have caused mixed reactions. They were valued for confirmation, enjoyment, and immersion in outdoor settings, but were criticized when mismatches occurred or when they sounded artificial. For example, P9 described sound effects as \textit{``double confirmation''} (e.g., boat sounds validating water scenes), P11 noted \textit{``pleasant''} effects in nature scenes (e.g., birds, wind) enhancing relaxation, and P4 mentioned that with sound effects, \textit{``nature scenes feel immersive''}. However, P4 recalled a \textit{``jarring''} instance where a rooster sound played for a red panda image. Multiple participants (P7, P11) requested spatialized and directional sound to locate objects (e.g., \textit{``traffic moving left to right''}). 


\noindent \textbf{3. Context-Dependent Usage Patterns:} 
Preferences could be described along two axes: indoor/outdoor and leisure/utility. Outdoors, sound effects are found to be immersive (P4: \textit{``wind sounds made me feel present,''}), whereas, indoors, they interfere with focus. Participants advocated for distinct modes: a \textit{``relaxation mode''} with rich audio for outdoor leisurely scenes such as parks, and a \textit{``utility mode''} with minimal sound effects for tasks such as reading signage. P2 also suggested safety applications, such as accident-prevention sounds (e.g., screeching tires) for hazard alerts.

\noindent\textbf{4. Usability Improvements and Hardware Proposals:} Although the app was praised for its intuitiveness, three areas of improvement dominate the feedback. First, latency reduction is requested by all. For example, P7 mentioned delays in bus number recognition that make the feature \textit{``useless in real-time.''} Second, participants ask for features such as OCR (P9: \textit{``Read bus numbers or signs''}) and real-time video descriptions (P11: \textit{``Like Google Translate’s live text scanning''}). Third, hardware integration ideas surfaced, with P2 envisioning \textit{``sunglasses with built-in cameras and audio''} to replace phone-based operation. Accessibility suggestions such as dark mode (P9) and question-answering (P11) with the device or app were also highlighted.

\section{Discussion}

\subsection{Balancing Comprehension and Engagement}
The experience of Vista spaces by BLVs through spoken descriptions, especially for leisure purposes, is not as pleasant and immersive as the visual experience \cite{gupta2024imwut,bandukda2019understanding}. Our quantitative and qualitative findings in the current study suggest that combining verbal descriptions with non-verbal sounds generated from our Scene2Audio framework are preferred over verbal descriptions only. Both engagement and immersion are improved, while remaining as comprehensible as speech only. 
 
This aligns with previous studies showing that nonverbal sounds help to build mental images \cite{rodero2020audio, bolls2003saw, steinhaeusser2021comparing, rodero2012see}, evoke emotions \cite{steinhaeusser2021comparing}, and improve immersion \cite{green2000role,hattich2020hear} in storytelling and narrative building. In addition, we did not observe any significant difference in the cognitive load of the verbal and non-verbal audio types, which highlights that additional non-verbal sound effects complement rather than overwhelm. 

However, non-verbal audio alone is unable to clearly communicate the scene to the users. The role of the non-verbal audio is to bring one into the scene and set the stage, while the spoken description provides clear understanding of the scene. This is consistent with previous research showing that to convey clear meaning, verbal descriptions are necessary in addition to sound effects, such as the way sound effects and speech are blended in movies for enhanced audio descriptions for BLVs \cite{crisell1994understanding,fryer2010audio,lopez2021enhancing}. We also found that verbal descriptions are important to clarify meanings because different sounds may have different meanings in different contexts or regions. For example, the traffic light beeping sound pattern differs between countries. Those beeps, when presented without context, could be interpreted in different ways. In fact, some participants mistook those traffic light beeps as tram bells. These findings are consistent with previous work that highlight the significant role that speech plays in disambiguating non-verbal sounds. 

\vspace{-0.1cm}
\subsection{Sonifying Scenes with Generative Models}
Insights from the Scene2Audio framework and the listening tests provide a valuable direction for designing end-to-end generative models that produce meaningful audio for Vista space description. Our findings show that identifying sonic elements, controlling the number of discrete events, and effectively managing their layering to represent a Vista space scene (the key components of the Scene2Audio framework) results in better  comprehension of the scene and improved pleasantness of the audio. This approach ensures that key elements are captured and their interactions (e.g., layering and event timing) do not overwhelm the listener. This is consistent with the principles in psycho-acoustics and auditory scene analysis \cite{moore2012introduction,bregman1994auditory} that human auditory system tends to easily capture the essence of a scene through auditory cues and the strategic organization of sounds. Baseline methods, on the other hand, struggled to capture and organize the salient objects especially in high density complex scenes, resulting in a less comprehensible and pleasant audio (Section \ref{sec:study1_results}). 
This highlights that existing generative models cannot be applied directly as plug-and-play solutions; instead, they require careful adaptation to suit the specific context and use case. 

Our approach aligns with the design principles recommended for generative AI applications \cite{weisz2024design}. In our framework design, we applied the `principle of designing for mental models', ensuring that the generated auditory scenes are consistent with how the human auditory system processes and comprehends sound. 

\subsection{Ecological Validity: Bridging Lab and In-the-Wild Findings}
Our comparative analysis shows interesting differences between the controlled lab study and the real-world app usage study, which can help to  define design principles for auditory scene description devices. The lab results show clear advantages of combining verbal descriptions with non-verbal sound effects for immersion and enjoyment, aligning with established literature on auditory scene analysis \cite{rodero2020audio, green2000role}. The in-the-wild results show how environmental and behavioral contexts affect these outcomes. An interesting difference emerged in the preference for the audio mode. Although lab participants consistently favored the overlay mode for leisure-oriented scenes because of its brevity and low cognitive load, mobile app users preferred the detail mode (overlay-concat) despite additional cognitive load. In fact, the lab study was a simulated environment, where the participants had to imagine the scene based on the sounds they heard. So a lot of details in the overlay-concat mode caused mental overload. In the wild, the participants already had a partial knowledge and understanding of the scene, and they were looking for additional details that the overlay-concat mode can provide. This behavioral shift reflects the role of environmental context in auditory attention. 
Human auditory attention is theorized to be goal-directed~\cite{kaya2017modelling}, i.e.~attention modulates auditory processing according to task demand. This theory can explain our finding that lab participants - who were imagining scenes without any context - tend to prefer the overlay mode that provides better mental imagery with decreased mental effort (task: ``imagining the scene''). In contrast, mobile app users, operating in real-world environments with partial scene understanding, sought additional details. In that case, the overlay-concat mode can improve their understanding of the scene or resolve ambiguities (task: ``better understanding the scene''). 

The environmental context also pointed to hallucinations of audio generative models. While well-matched naturalistic sounds like P4’s \textit{``snow footsteps''} enhanced immersion, mismatch between sound effect and verbal description (for example, P4's \textit{``rooster for red panda''}) led to cognitive dissonance. Such hallucinations are expected in audio generative models, especially when sounds of such type are rare in audio databases that they are trained on.


Finally, several practical constraints emerged in the mobile-app study. For instance, the latency of the framework in generating audio did not affect much in leisure scenes, which were typically Vista space scene. However, for non-leisure non-Vista space scenes, the users mentioned the inability to use the app, especially in time-sensitive tasks (P7: ``Bus info arrived too late''). Another practical constraint was that ambient noise, especially outdoors, frequently rendered subtle sounds  inaudible (P9: ``Faint sounds were unusable outdoors''). These findings suggest lab studies best evaluate perceptual potential, whereas in-the-wild testing exposes operational thresholds for deployment. 

\newtext{Although participants occasionally used the system to obtain functional or safety-related information (e.g., locating paths, identifying slopes, or determining nearby structures), our current prototype is not designed or validated for such utility use. Given the presence of occasional hallucinations, the system may produce incomplete or misleading outputs in real-world contexts. For responsible deployment, we explicitly caution that the current implementation should not be used for navigation, hazard awareness, or decision-making tasks. Instead, it should be treated as a relaxation- and exploration-oriented tool that provides an experiential rendering of scenes. Trustworthiness of AI is an active area of research and future work must incorporate rigorous safety evaluations, and  hallucination mitigation for better trustworthiness of the system regarding mobility.}


\section{Limitations \& Future Work}

\paragraph{\textbf{Limited number of BLV participants}}
The study was conducted with a small number of BLV participants due to limited access to this population, which may limit the generalization of the findings. Subjective assessments with a low number of participants to draw design implications is considered a reliable method \cite{nielson2012,pernice2021}, especially when working with a niche population such as BLV \cite{brule2020review,aziz2019investigation,sears2011representing,boldu2020aisee,constantinescu2020bring}. The participants in our study had a diverse range of visual conditions, which helped us capture a variety of perspectives within the limitations of a smaller sample size. In addition, the consistency in the feedback from the participants provides valuable insights for future research in this area. 

\paragraph{\textbf{Controllable Gen-AI for Scene Sonification}}
Ensuring high-quality, controllable sound generation remains a challenge. Our findings highlight the need of controlling the number of discrete events in the generated audio. Our method of generating multiple samples and selecting the one with the fewest discrete events offers a partial solution. A more seamless control over features of the generated sound that is integrated with the AI model is still needed. As there is currently no straightforward method for controlling the discrete events in the generated audio \cite{kamath2024morphfader,xie2024picoaudio,kamath2024example, gupta2023towards}
, future research should focus on improving this capability in generative models to ensure more precise sound generation. \newtext{Additionally, our mixing strategy used fixed attenuation values rather than perceptual loudness targets. Incorporating LUFS-based normalization (e.g., EBU R128\footnote{\url{https://en.wikipedia.org/wiki/EBU_R_128}}) would improve perceptual consistency across scenes and robustness in real-world listening conditions.} \newtext{For speech descriptions, we used a single TTS voice with fixed prosody for all conditions to avoid confounds. 
Future work can explore controllable TTS prosody and personalised voice settings.}
\paragraph{\textbf{Practical Deployment Considerations}}
Our in-the-wild study reveals three practical challenges for deployment: 
(1) the need for context-based adaptive audio systems or a flexible modality control interface that can toggle between utility mode and leisure-focused modes, for example, disabling sound effects indoors or in a utility-focused context, while providing the sound effects outdoors or in a leisure-focused context, (2) reducing latency for mobility-related use-cases, and (3) integrating with a wearable device to address hands-free interactions.  Future systems should incorporate context-aware generative models, and low-latency pipelines to bridge the gap between lab-based scenes and dynamic everyday
environments.

\newtext{We acknowledge that spatialization and distance cues play an important role in situated awareness and real-world immersion, and we view them as complementary, rather than alternative, to our approach. Our framework is technically compatible with spatial audio engines, and integrating spatialization constitutes a natural next step. Future work will investigate how spatial cues interact with the generated scene content, alongside appropriate validation, particularly for outdoor navigation and orientation scenarios.}

\paragraph{\textbf{Scene-Level Considerations}}
\newtext{Some vista scenes naturally contain minimal sonic content (e.g., stars, clouds). In such cases, our system defaults to speech-only descriptions rather than synthesising artificial sounds. We also acknowledge the challenge of misaligned photos in blind photography \cite{jayant2011supporting}; however, this issue is orthogonal to our aim of evaluating the perceptual qualities of scene-to-audio mappings. Future deployments can integrate established camera-guidance methods to improve input framing.} \newtext{Finally, our system captures the “Vista gist”, i.e.~the cognitive snapshot a sighted person forms when pausing to appreciate a distant view from a vantage point. Integrating dynamic spatial cues and multimodal exploration interfaces would extend the framework toward navigational applications and richer spatial awareness.} 
\newtext{A related limitation concerns participants’ familiarity with the instructed vantage points (hilltop or rooftop). 
We did not measure prior experience with such contexts, and this could influence the ease or vividness of mental imagery, particularly for BLV participants who may not have encountered these viewpoints. 
However, all participants evaluated the same scenes under identical instructions, making systematic bias across audio conditions unlikely. 
An interesting direction that emerges from our findings is how users’ familiarity with specific vista-like vantage points may shape how easily they imagine or interpret distant-scene audio.
}

\section{Conclusion}
In this work, we design Scene2Audio framework to convert Vista spaces into representative non-verbal sounds. 
    Our 
    framework significantly outperforms baselines like Im2wav generating a representative and pleasant auditory experience for a given scene image, especially for nature scenes. 
    The Overlay method, which simultaneously combines speech and non-verbal sounds, provides the most effective balance of comprehension, quality, immersion, and cognitive load. This approach is particularly beneficial for BLV, offering a richer and more accessible understanding of Vista spaces, taking a step towards bridging the gap between visual and auditory experience.
    
We believe these findings offer significant value to the CHI community on how visual information in Vista spaces can be effectively conveyed through auditory means.  
Finally, our work could potentially influence technologies beyond the specific application scenario considered in this paper, such as visual media content like movies, virtual and augmented reality, where automatically generating comprehensible, enjoyable, and immersive auditory cues to represent visual scenes play an important role.
\begin{acks}
We would like to thank the BLV participants and Singapore Association of the Visually Handicapped (SAVH) for their time and support for this work.
\end{acks}
\bibliographystyle{ACM-Reference-Format}
\bibliography{sample-base}

\appendix
\newpage
\section*{Appendix}
\begin{enumerate}
    \item \textbf{Comparison of sounds generated from Scene-to-Audio framework with through generated from baseline techniques}
    \begin{table}[h!]
\centering
\begin{minipage}{0.5\textwidth}
\centering
\caption{Comparison of our framework with baseline methods in scene selection. All numbers are in \% accuracy in selecting the correct scene image that was used to generate that sound, out of four scene image options. Each sound-type for every scene-type was evaluated by a total of 21 participants. Pairwise statistical significance using Wilcoxon signed rank test is shown as superscript of the corresponding audio type on the higher score (p<0.05). A superscript with * means p<0.005.}
\label{table:comparison}
\vspace{-0.3cm}
\resizebox{\columnwidth}{!}{
\begin{tabular}{|c|c|c|c|}
\hline
  \textbf{Scene}      & \textbf{Im2wav$^{(a)}$} & \textbf{Im2text2audio$^{(b)}$} & \textbf{Ours$^{(c)}$} \\ \hline
seabeach & 0     & \textbf{95$^{(a*)}$}            & 76$^{(a*)}$   \\ \hline
mountains & 0     & 29$^{(a)}$            & \textbf{62$^{(a*,b)}$}   \\ \hline
countryside & 0     & 14            & \textbf{81$^{(a*,b*)}$}   \\ \hline
reservoir & 24$^{(b)}$    & 0             & \textbf{57$^{(b)}$}   \\ \hline
park      & 5     & 14            & \textbf{90$^{(a*,b*)}$}   \\ \hline
train station & \textbf{33}    & 14            & \textbf{33}   \\ \hline
foodcourt & 5     & 57$^{(a*)}$            & \textbf{76$^{(a*)}$}   \\ \hline
street    & \textbf{71$^{(b*,c)}$}    & 14            & 24   \\ \hline
AVERAGE   & 17.25 & 29.63$^{(a)}$         & \textbf{62.38$^{(a*,b*)}$} \\ \hline
\end{tabular}}
\end{minipage}
\end{table}
\item \textbf{Details of the Questionnaire for the User Evaluation Study with BLVs}
\subsubsection*{Comprehension}
Comprehension evaluates how well the sonified scenes convey information to the listener. We measured both subjective understanding and objective accuracy of the scene comprehension with two questions:

\begin{itemize}
    \item An open-ended question: \textit{``Can you describe the scene that you just heard?''}\\
Participants' descriptions were analyzed using an Intent Similarity score, which calculates cosine similarity between the word embedding of their description and a reference description generated by ChatGPT (the text used in Speech-only audio)\footnote{The word embeddings were extracted using python's NLP library spacy's English language model en\_core\_web\_sm}. The score was scaled to the range 1 to 7 for comparison with other questions.

\item Likert scale rating for the statement: \textit{``I could comprehend the scene from this experience.''}\\
This 7-point scale allowed participants to rate their own understanding of the scene, providing a subjective measure of comprehension.

\end{itemize}

\subsubsection*{Engagement}
We assess the quality of the auditory engagement across four dimensions: enjoyment, curiosity, imagination, and memorability. These dimensions are informed by HCI, psychology, and auditory design frameworks. Participants rated the following statements on a 7-point Likert scale:

\begin{itemize}
    \item \textbf{Enjoyable:} \textit{``This experience was enjoyable.''}\\
    We use this dimension because according to Flow Theory \cite{czikszentmihalyi1990flow},  
    when users find an experience enjoyable, they are more likely to be fully absorbed in it, leading to a deeper engagement with the content.
    \item \textbf{Curiosity:} \textit{``This experience made me curious.''}\\
    By evaluating curiosity, we assess whether the auditory feedback stimulates participants' desire to explore and understand the scene further \cite{berlyne1954theory}, indicating a higher quality of engagement.
    \item \textbf{Imagination:} \textit{``This experience aroused my imagination.''}\\
    We evaluate this dimension as %
    sounds have the potential to stimulate imagination and create vivid mental representations of the scene \cite{steinhaeusser2021comparing, rodero2012see}, that enhances engagement.
    \item \textbf{Memorability:}
    \textit{``This experience was memorable.''}\\
    We evaluate this dimension because memorability of experiences is related to engagement of emotions and imagination  
    \cite{rodero2019spark, potter2006made}. 
\end{itemize}

\subsubsection*{Immersion}
Immersion measures how deeply participants feel absorbed in the auditory experience. 
We assess immersion of the audio content without spatializing sounds. We use Igroup Presence Questionnaire (IPQ) \cite{melo2023IPQ}, which is a standard self-assessment questionnaire for evaluating how immersed or ``present'' a person feels within a virtual environment. It is a 7 point Likert scale for rating the following statements, each linked to a factor for immersion provided in IPQ:
\begin{itemize}
    \item Presence: \textit{``In this experience, I had a sense of being present in the scene I was listening to.''}
    \item Spatial Presence: \textit{``I felt that the scene surrounded me.''}
    \item Involvement: \textit{``During this experience, I was not aware of my real environment.''}
    \item Experienced Realism: \textit{``The experience seemed like the real world to me.''}
\end{itemize}
\subsubsection*{Cognitive Load}
We used the NASA-TLX (Task Load Index) questionnaire \cite{hart1988development} to measure the cognitive effort required to process the auditory information. This tool evaluates cognitive load across various dimensions, including mental demand, physical demand, and perceived success, ensuring that the experience is accessible and not overly taxing for the listener.

\textbf{Additional Questions:}
In addition to the questions related to the four criteria, we asked some general questions at the end of listening to the eight audio clips.  

\subsubsection*{Preference Rank-Order}
We asked the participants to provide a rank-order of their preference of the 4 audio types they heard.

\subsubsection*{Open-Ended Questions}
\label{sec:openquestions}

\begin{itemize}
    \item Why do you prefer the audio type that you ranked first? 
    
    \item In your opinion, what is the role of speech and what is the role of non-verbal audio in describing scenes? 
     
\end{itemize}
\item \textbf{Detailed Ratings for Different Audio Types in the User Study with BLVs}
\begin{table*}
\centering
\begin{minipage}{0.45\textwidth}
\centering
\includegraphics[width=\textwidth]{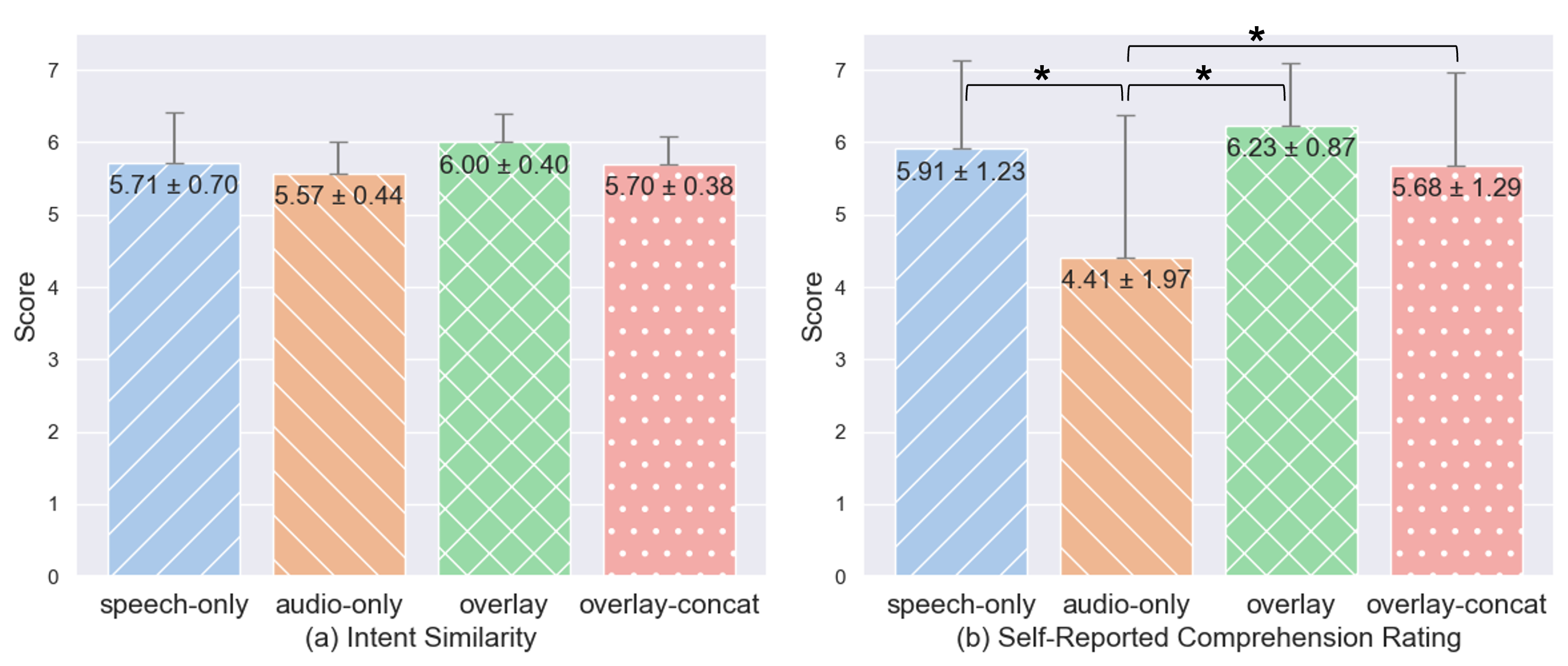}
\vspace{-0.7cm}
\captionof{figure}{Ratings for the Comprehension Criteria Questions.
} 
\Description{Two side-by-side bar charts compare four conditions—speech-only, audio-only, overlay, and overlay-concat—on comprehension-related ratings.
The left chart (intent similarity) shows similar scores across conditions, with overlay highest (about 6.0), followed by speech-only and overlay-concat (about 5.7), and audio-only slightly lower (about 5.6).
The right chart (self-reported comprehension) shows larger differences: overlay is highest (about 6.2), speech-only and overlay-concat are similar (about 5.7–5.9), and audio-only is substantially lower (about 4.4) with higher variability. Statistical significance markers indicate overlay is significantly higher than audio-only and other comparisons.}
\label{fig:comprehension}
\end{minipage}
\begin{minipage}{0.45\textwidth}
\centering
\includegraphics[width=\textwidth]{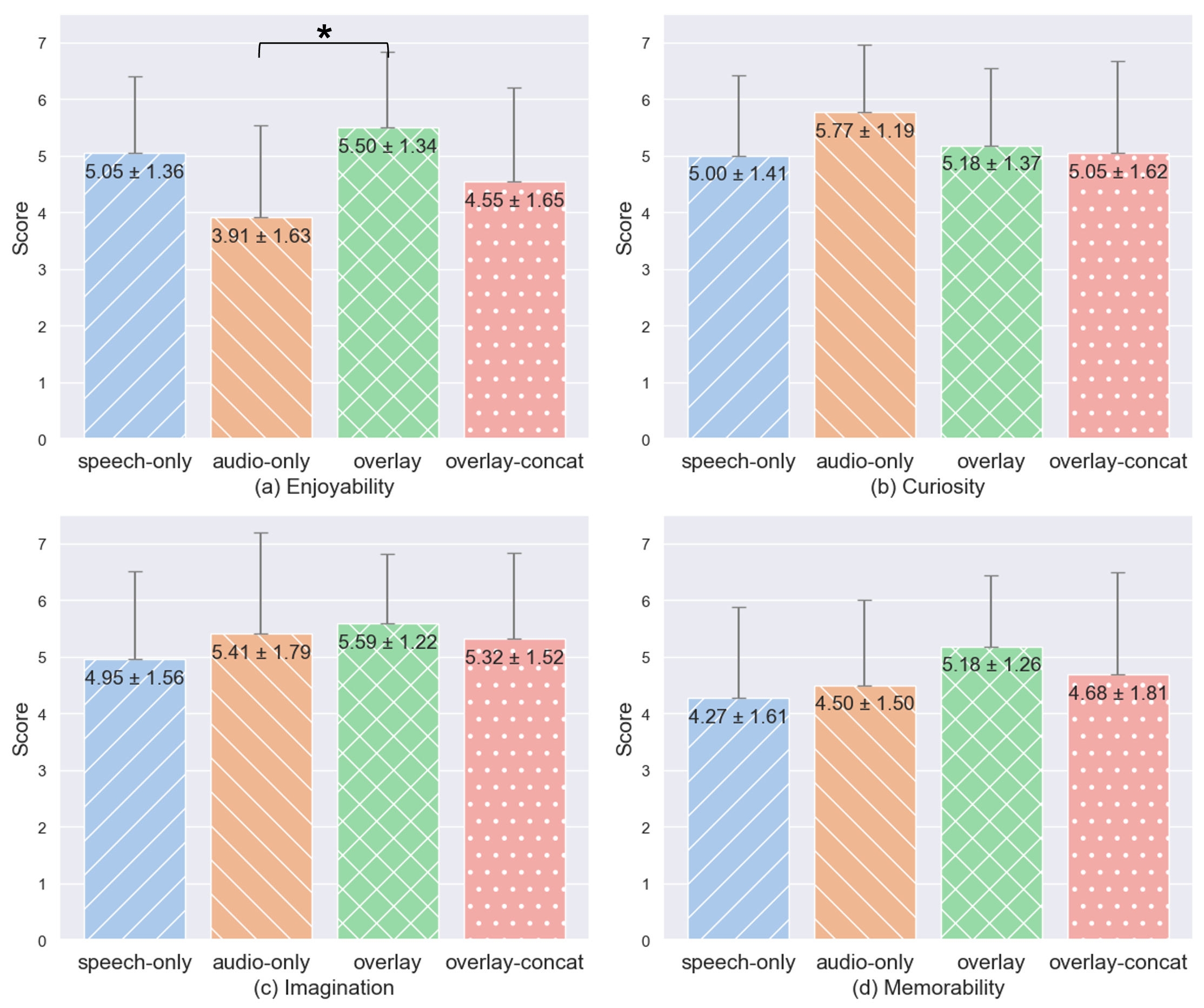}
\vspace{-0.7cm}
\captionof{figure}{Ratings for the Engagement Criteria Questions.
}
\Description{Four bar charts compare four conditions—speech-only, audio-only, overlay, and overlay-concat—across engagement measures: enjoyability, curiosity, imagination, and memorability.
In enjoyability, overlay has the highest score (about 5.5), followed by speech-only (about 5.0) and overlay-concat (about 4.6), while audio-only is lowest (about 3.9); a significance marker indicates overlay is significantly higher than audio-only.
In curiosity, audio-only is highest (about 5.8), with overlay, overlay-concat, and speech-only clustered slightly lower (about 5.0–5.2).
In imagination, overlay is highest (about 5.6), followed closely by audio-only and overlay-concat (about 5.3–5.4), with speech-only slightly lower (about 5.0).
In memorability, overlay again scores highest (about 5.2), followed by overlay-concat and audio-only (about 4.6–4.7), and speech-only lowest (about 4.3). Overall, overlay consistently performs best across most engagement measures, while audio-only shows mixed performance.}
\label{fig:engagement}
\end{minipage}
\end{table*}

\begin{table*}
\centering
\begin{minipage}{0.48\textwidth}
\centering
\includegraphics[width=\textwidth]{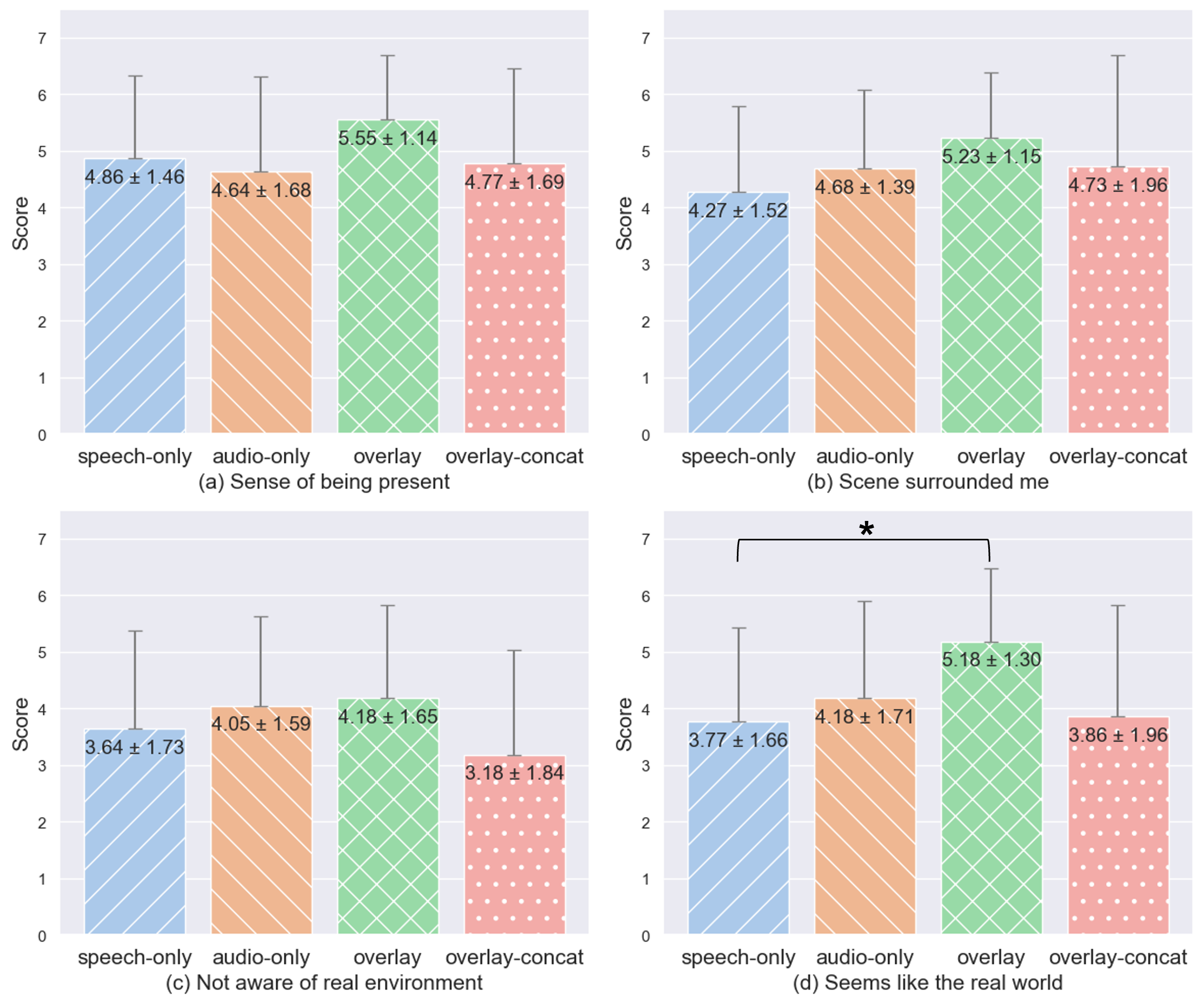}
\vspace{-0.7cm}
\captionof{figure}{Ratings for the Immersion Criteria Questions.
} 
\Description{Four bar charts compare four conditions—speech-only, audio-only, overlay, and overlay-concat—across immersion measures: sense of presence, feeling surrounded by the scene, reduced awareness of the real environment, and realism.
Overlay consistently scores highest across all measures (about 5.2–5.6), indicating stronger immersion. Speech-only and audio-only show moderate scores (about 4.2–4.9), while overlay-concat is similar or slightly lower.
For “seems like the real world,” overlay is significantly higher (about 5.2) than speech-only (about 3.8). Overall, overlay provides the strongest immersive experience, while other conditions cluster at lower levels.}
\label{fig:immersionquestions}
\end{minipage}
\begin{minipage}{0.48\textwidth}
\centering
\includegraphics[width=\textwidth]{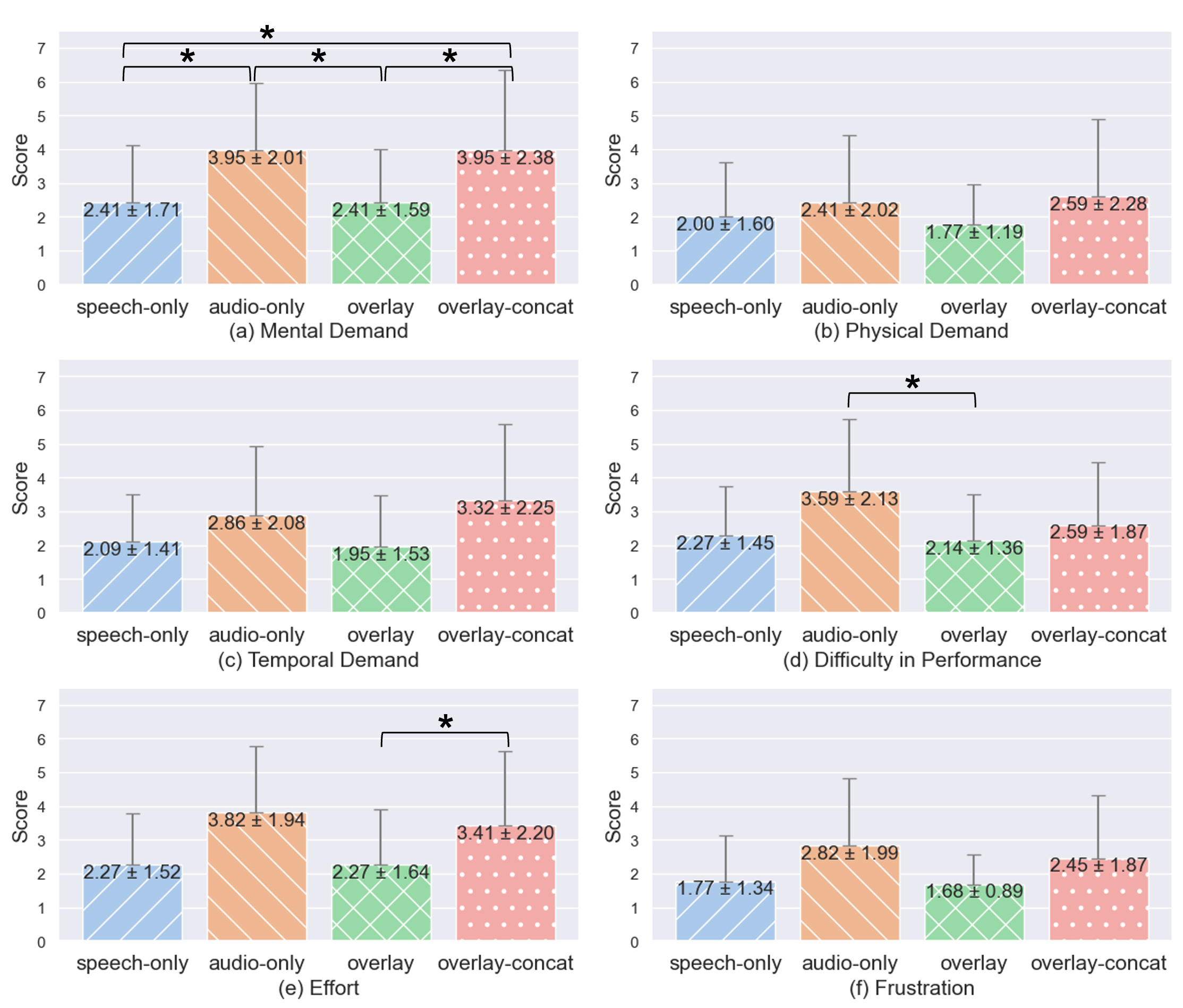}
\vspace{-0.7cm}
\captionof{figure}{Ratings for the Cognitive Load Criteria Questions.
}
\Description{Six bar charts compare four conditions—speech-only, audio-only, overlay, and overlay-concat—across cognitive load measures: mental, physical, and temporal demand, difficulty, effort, and frustration.
Audio-only and overlay-concat generally show higher cognitive load, while overlay and speech-only are lower. In mental demand, audio-only and overlay-concat are highest (about 4.0), while speech-only and overlay are lower (about 2.4), with significant differences indicated.
Physical and temporal demand show smaller differences, though audio-only and overlay-concat trend higher. In difficulty, audio-only is significantly higher (about 3.6) than overlay (about 2.1).
Effort is also higher for audio-only and overlay-concat (about 3.4–3.8) compared to speech-only and overlay (about 2.3). Frustration follows a similar pattern, with audio-only highest and overlay lowest.
Overall, overlay and speech-only result in lower cognitive load, while audio-only and overlay-concat impose higher demand.}
\label{fig:cognitiveload}
\end{minipage}
\end{table*}

\begin{table*}
\caption{Comparison of Ratings for Different Audio Types. Pairwise statistical significance using Wilcoxon signed rank test is shown as superscript of the corresponding audio type on the higher score (p<0.05). A superscript with * means p<0.005. For the four criteria, $\uparrow$ means higher is better and $\downarrow$ means lower is better.}
\label{tab:audio_scores}
\vspace{-0.3cm}
\centering
\resizebox{\textwidth}{!}{
\begin{tabular}{|l|l|c|c|c|c|}
\hline
\textbf{Criteria} & \textbf{Questions} & \textbf{Speech-only$^{(a)}$} & \textbf{Audio-only$^{(b)}$} & \textbf{Overlay$^{(c)}$} & \textbf{Overlay Concat$^{(d)}$} \\ \hline
\multirow{2}{*}{Comprehension $\uparrow$} & Intent Similarity & 5.71$\pm$0.70 & 5.57$\pm$0.44 & \textbf{6.00$\pm$0.40}$^{(b,d)}$ & 5.70$\pm$0.38 \\ \cline{2-6} 
 & Comp Rating & 5.91$\pm$1.23$^{(b)}$ & 4.41$\pm$1.97 & \textbf{6.23$\pm$0.87}$^{(b*)}$ & 5.68$\pm$1.29$^{(b)}$ \\ \hline
\multirow{4}{*}{Qual. of Experience $\uparrow$} & Enjoyable & 5.05 $\pm$ 1.36$^{(b)}$ & 3.91 $\pm$ 1.63 & \textbf{5.50 $\pm$ 1.34}$^{(b*,d)}$ & 4.55 $\pm$ 1.65 \\ \cline{2-6} 
 & Curious & 5.00 $\pm$ 1.41 & \textbf{5.77 $\pm$ 1.19}$^{(a)}$ & 5.18 $\pm$ 1.37 & 5.05 $\pm$ 1.62 \\ \cline{2-6} 
 & Imagination & 4.95 $\pm$ 1.56 & 5.41 $\pm$ 1.79 & \textbf{5.59 $\pm$ 1.22} & 5.32 $\pm$ 1.52 \\ \cline{2-6} 
 & Memorable & 4.27 $\pm$ 1.61 & 4.50 $\pm$ 1.50 & \textbf{5.18 $\pm$ 1.26}$^{(a)}$ & 4.68 $\pm$ 1.81 \\ \hline
\multirow{4}{*}{Immersion $\uparrow$} & Presence & 4.86 $\pm$ 1.46 & 4.64 $\pm$ 1.68 & \textbf{5.55 $\pm$ 1.14} & 4.77 $\pm$ 1.69 \\ \cline{2-6} 
 & Spatial presence & 4.27 $\pm$ 1.52 & 4.68 $\pm$ 1.39 & \textbf{5.23 $\pm$ 1.15}$^{(a)}$ & 4.73 $\pm$ 1.96 \\ \cline{2-6} 
 & Involvement & 3.64 $\pm$ 1.73 & 4.05 $\pm$ 1.59 & \textbf{4.18 $\pm$ 1.65}$^{(d)}$ & 3.18 $\pm$ 1.84 \\ \cline{2-6} 
 & Experienced realism & 3.77 $\pm$ 1.66 & 4.18 $\pm$ 1.71 & \textbf{5.18 $\pm$ 1.30}$^{(a*,b,d)}$ & 3.86 $\pm$ 1.96 \\ \hline
\multirow{6}{*}{Cognitive Load $\downarrow$} & Mental Demand & \textbf{2.41 $\pm$ 1.71}$^{(b,d)}$ & 3.95 $\pm$ 2.01 & \textbf{2.41 $\pm$ 1.59}$^{(b,d)}$ & 3.95 $\pm$ 2.38 \\ \cline{2-6} 
 & Physical Demand & 2.00 $\pm$ 1.60 & 2.41 $\pm$ 2.02 & \textbf{1.77 $\pm$ 1.19}$^{(d)}$ & 2.59 $\pm$ 2.28 \\ \cline{2-6} 
 & Temporal Demand & 2.09 $\pm$ 1.41$^{(d)}$ & 2.86 $\pm$ 2.08 & \textbf{1.95 $\pm$ 1.53}$^{(d*)}$ & 3.32 $\pm$ 2.25 \\ \cline{2-6} 
 & Performance (inverted) & 2.27 $\pm$ 1.45$^{(b)}$ & 3.59 $\pm$ 2.13 & \textbf{2.14 $\pm$ 1.36}$^{(b)}$ & 2.59 $\pm$ 1.87$^{(b)}$ \\ \cline{2-6} 
 & Effort & \textbf{2.27 $\pm$ 1.52 }$^{(b,d)}$& 3.82 $\pm$ 1.94 & \textbf{2.27 $\pm$ 1.64}$^{(b,d)}$ & 3.41 $\pm$ 2.20 \\ \cline{2-6} 
 & Frustration & 1.77 $\pm$ 1.34 & 2.82 $\pm$ 1.99 & \textbf{1.68 $\pm$ 0.89}$^{(b)}$ & 2.45 $\pm$ 1.87 \\ \hline
\end{tabular}}
\end{table*}

\pagebreak
\item \textbf{Questionnaires in the Mobile App Study}
See Figures \ref{fig:inappques1} and \ref{fig:inappques}
\begin{figure*}
    \centering
    \begin{subfigure}{0.32\textwidth}
        \includegraphics[width=\linewidth]{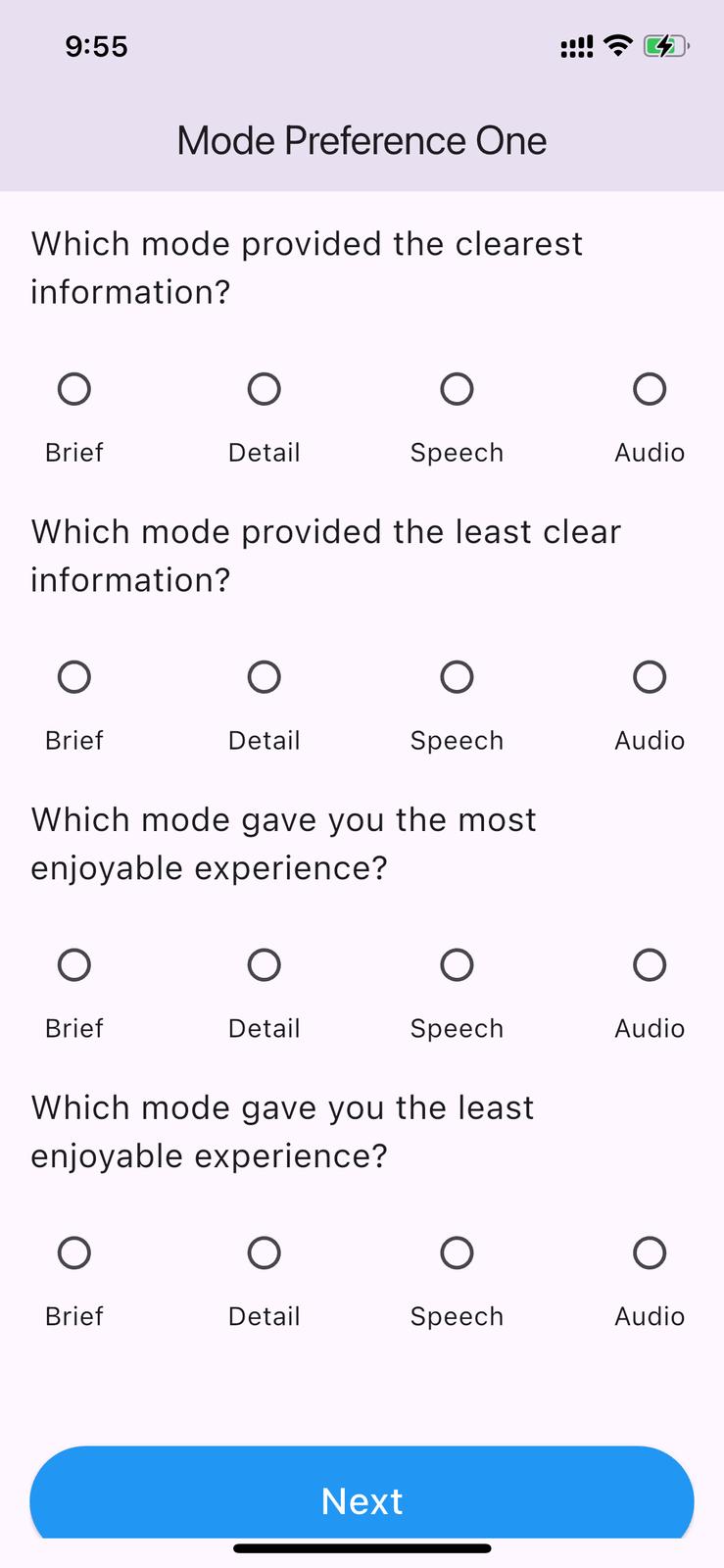}
    \end{subfigure}
    \begin{subfigure}{0.32\textwidth}
        \includegraphics[width=\linewidth]{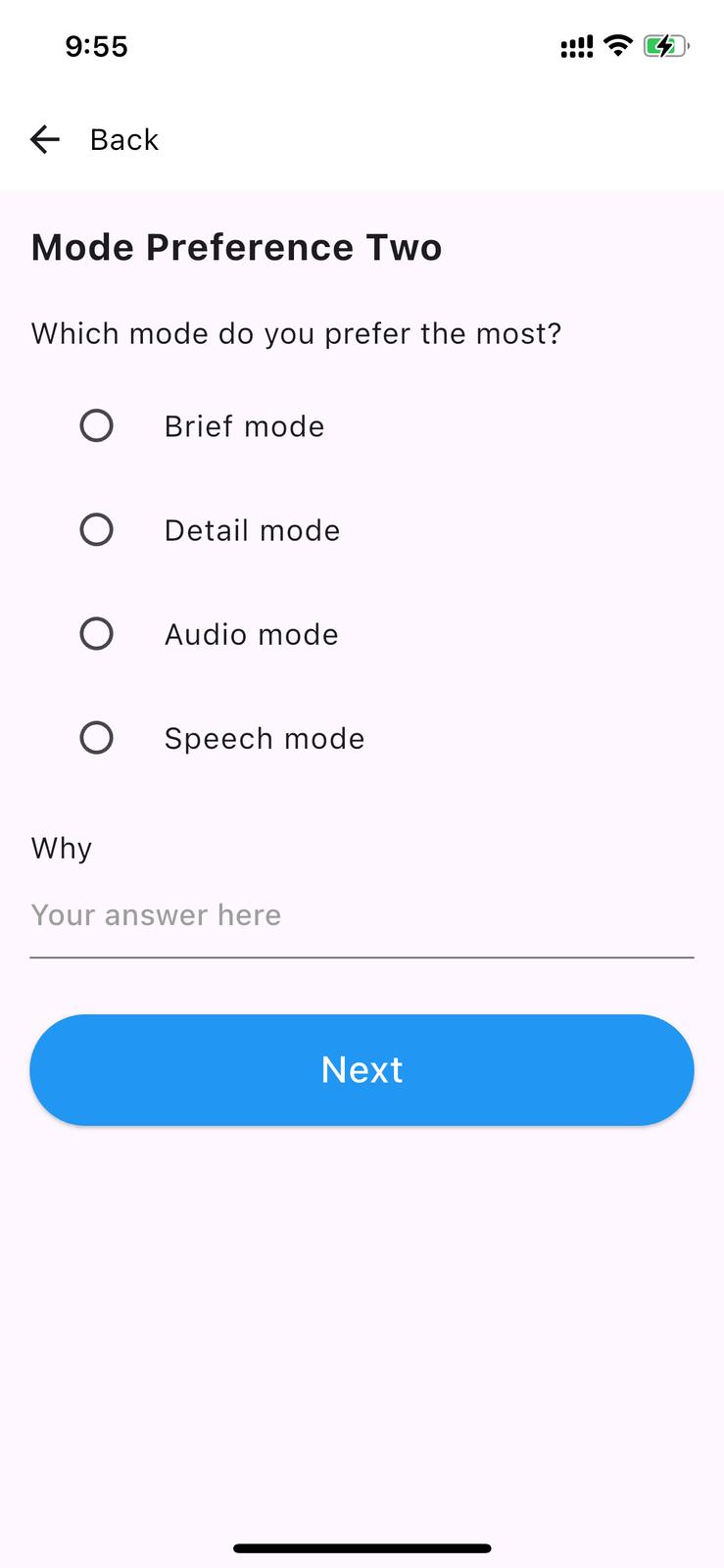}
    \end{subfigure}
    \begin{subfigure}{0.32\textwidth}
        \includegraphics[width=\linewidth]{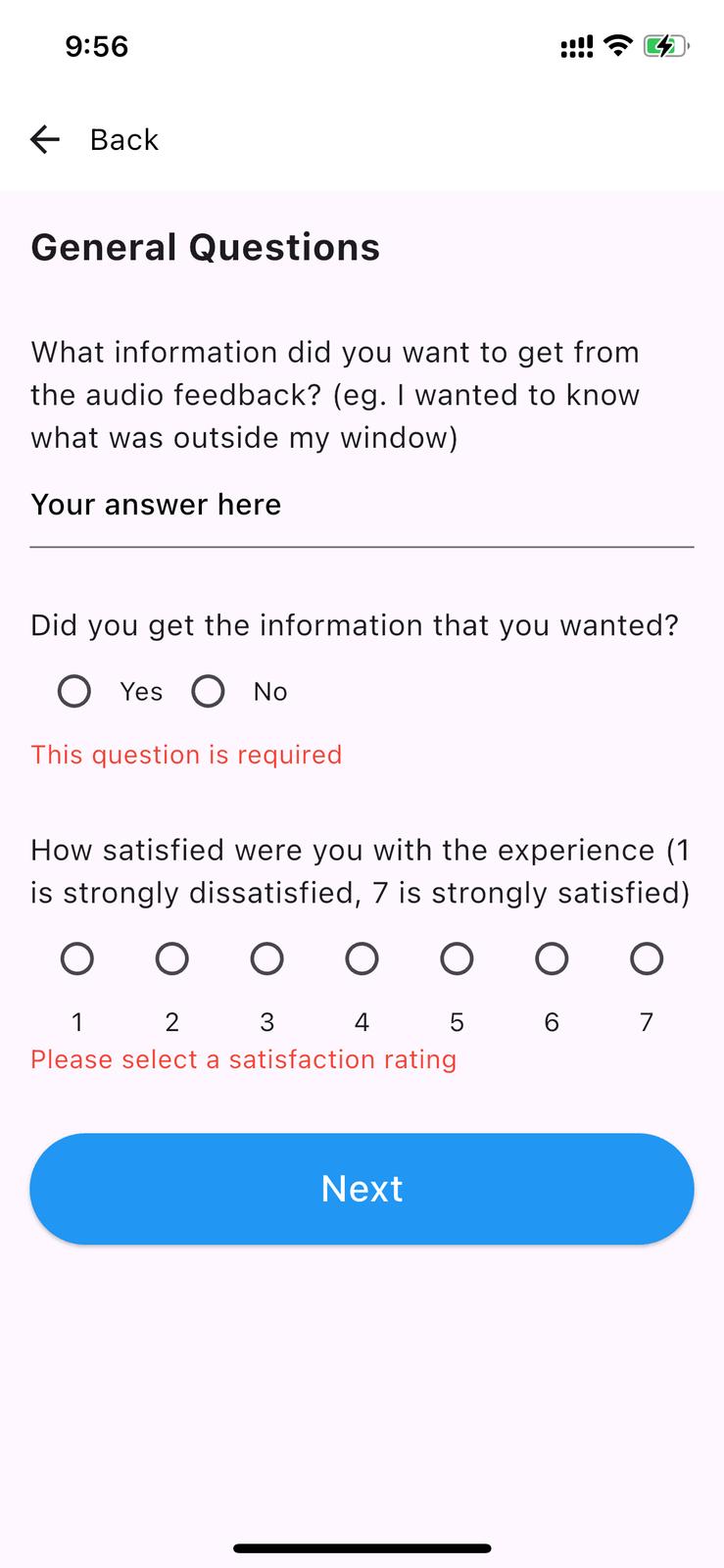}
    \end{subfigure}
    \caption{In-app questionnaire after each photo submission and audition of the sounds.}
    \label{fig:inappques1}
    \Description{Three mobile app screens show the post-task questionnaire used after photo submission and listening to generated audio.
The first screen (“Mode Preference One”) asks users to select which mode—brief, detail, speech, or audio—provided the clearest and least clear information, and the most and least enjoyable experience, using radio buttons.
The second screen (“Mode Preference Two”) asks users to select their most preferred mode (brief, detail, audio, or speech) and provide a short explanation.
The third screen (“General Questions”) includes an open-ended question about desired information, a yes/no question on whether the information was obtained, and a 1–7 satisfaction rating scale. Each screen includes a “Next” button for navigation.}
\end{figure*}

\begin{figure*}
    \centering
    \begin{subfigure}{0.32\textwidth}
        \includegraphics[width=\linewidth]{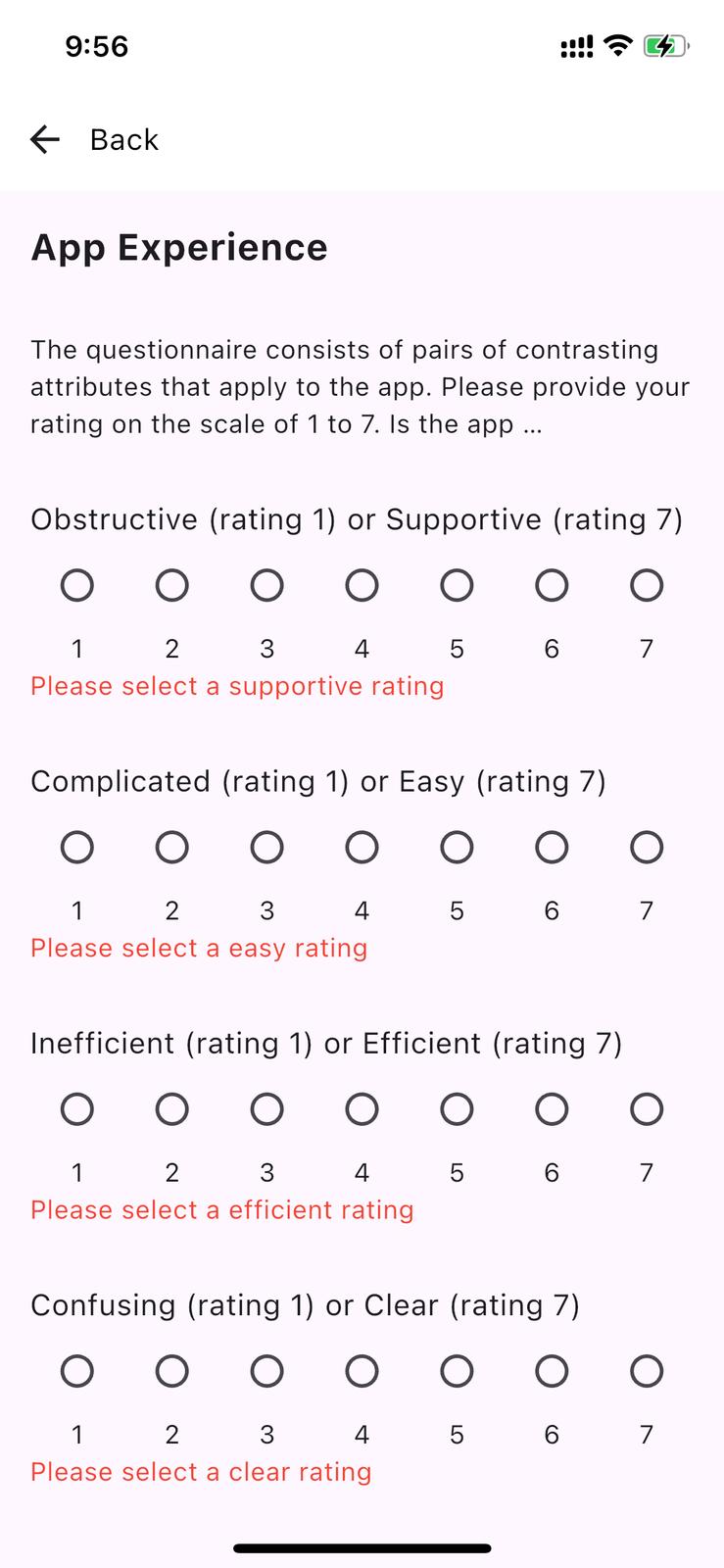}
    \end{subfigure}
    \begin{subfigure}{0.32\textwidth}
        \includegraphics[width=\linewidth]{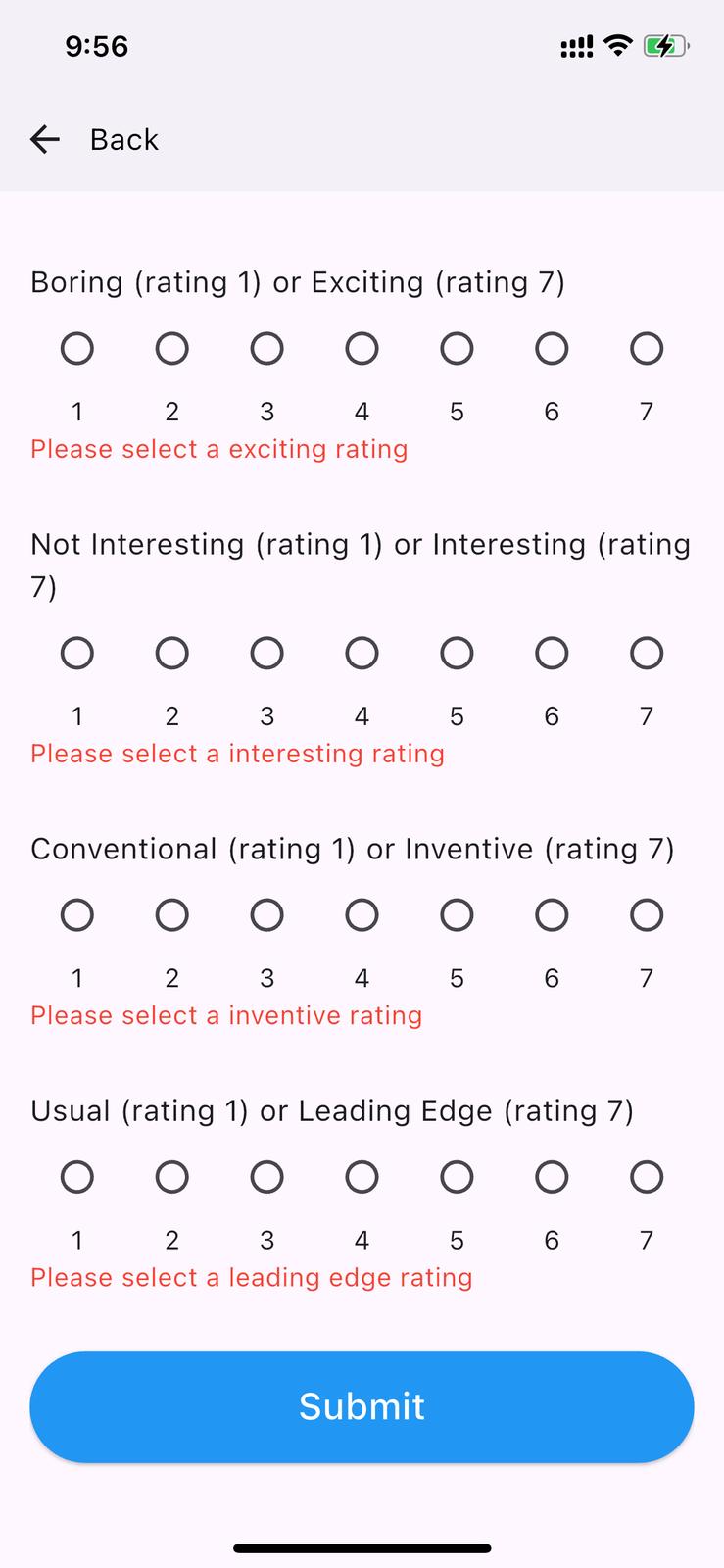}
    \end{subfigure}
    \caption{In-app questionnaire: User Experience Questionnaire (UEQ).}
    \Description{Two mobile app screens show an in-app questionnaire evaluating overall app experience after each task.
The questionnaire presents pairs of contrasting attributes rated on a 1–7 scale, including obstructive–supportive, complicated–easy, inefficient–efficient, confusing–clear, boring–exciting, not interesting–interesting, conventional–inventive, and usual–leading edge. Each item uses radio buttons with validation prompts if unanswered.
The final screen includes a “Submit” button to complete the questionnaire. Overall, the interface collects user ratings of usability, clarity, engagement, and perceived innovation.}
    \label{fig:inappques}
\end{figure*}

\item  \textbf{Evaluation of the Scene2Audio Framework}
\label{sec:proposed_eval}
We conducted a listening test with sighted participants to compare our Scene2Audio framework with two baseline methods. Our evaluation focused on participants' ability to match generated audio to corresponding images, as well as their subjective experience of the audio in terms of relevance and pleasantness.

\textbf{Baselines} 
We compared our framework with two baseline methods: 
\begin{itemize}
    \item \textbf{Im2wav \cite{sheffer2023hear}}: Directly maps images to audio, serving as a straightforward image-to-audio generation model.
    \item \textbf{Im2text2audio}: Uses a cascaded approach where images are first converted to textual descriptions via GPT-4, and then to audio using the AudioGen model \cite{kreuk2022audiogen}. Unlike our framework, Im2text2audio does not rely on enhancements based on psychoacoustics, auditory scene analysis, and Foley sound synthesis.
\end{itemize}
All models, including our framework, generated sounds that were 4-5 seconds long, as the models have been trained on audio clips of that length. These baselines provided benchmarks to highlight the improvements introduced by our Scene2Audio framework.

\textbf{Participants} The listening test involved 21 sighted participants (10 male, 11 female, average age 29 years, SD 7.5 years). 
In this study, our aim was to compare the sounds generated by the scene2audio framework with the baselines to sonify the visual elements of the scenes. This was a validation step to ensure that the visual elements are appropriately represented and that the generated audio is comprehensible. Hence, sighted participants were required for this study.

\textbf{Design of the Listening Test.} The test was carried out online using Qualtrics\footnote{\url{https://www.qualtrics.com/}}. The test consisted of 8 pages, each corresponded to a scene. Each page consisted of three sound clips, one from each method (our framework and the two baselines), generated for a single scene. As such, there were a total of 24 sound clips (8 scenes x 3 sound types) across 8 pages. The order of the scene was randomized and the order of the types of sound was counterbalanced to avoid sequence bias.

Alongside each audio file, four images were shown: one correct image (used to generate the audio) and three distractors. 
The distractor images had some overlapping sonic objects with the correct image but differed in the number or type of objects. 
Participants were instructed to listen carefully before selecting the image that most closely matched the audio. The study was approved by the Institutional Review Board (IRB).

The readers are encouraged to consult our website\footnote{\url{https://scenetoaudio.github.io/scenetoaudio/\#/section-3-4}} (anonymized for review) to listen to the audio clips and see the images used for this study.



\textbf{Evaluation Metrics}
We measured accuracy, defined as the percentage of times that participants correctly identified the image that generated the audio, i.e.~the ``correct'' image. This metric indicates how easy was it for the participants to identify the right scene image based on listening to the audio, which relates to comprehension of the scene based on the audio. Participants also rated the confidence in their image selection as well as pleasantness of the audio on a 7-point Likert scale. 

\textbf{Results and Discussion}
\label{sec:study1_results}
The results show the effectiveness of our Scene2Audio framework compared to the Im2wav and Im2text2audio baseline methods. As shown in Figure \ref{fig:comparison}, Scene2Audio outperforms the two baselines in 5 out of 8 scenes, particularly in nature environments such as mountains, countryside, and reservoir. It achieves an average accuracy of 62.4\% 
in selecting the correct scene image. A Wilcoxon signed rank test with Bonferroni correction shows that this is significantly higher than Im2text2audio's 29.6\% 
and Im2wav's 17.3\% 
(p<0.05). However, in urban scenes such as train stations, our framework performs comparably to Im2wav, and significantly lags in the street scene, achieving 24\% accuracy compared to Im2wav's 71\%. Such urban scenes often consist of inanimate objects, that result in ambiguous generated sounds that overlap with the distractor images and cause confusions (see qualitative feedback results below).

\begin{table*}
\centering
\begin{minipage}{0.65\textwidth}
\centering
\includegraphics[width=\textwidth]{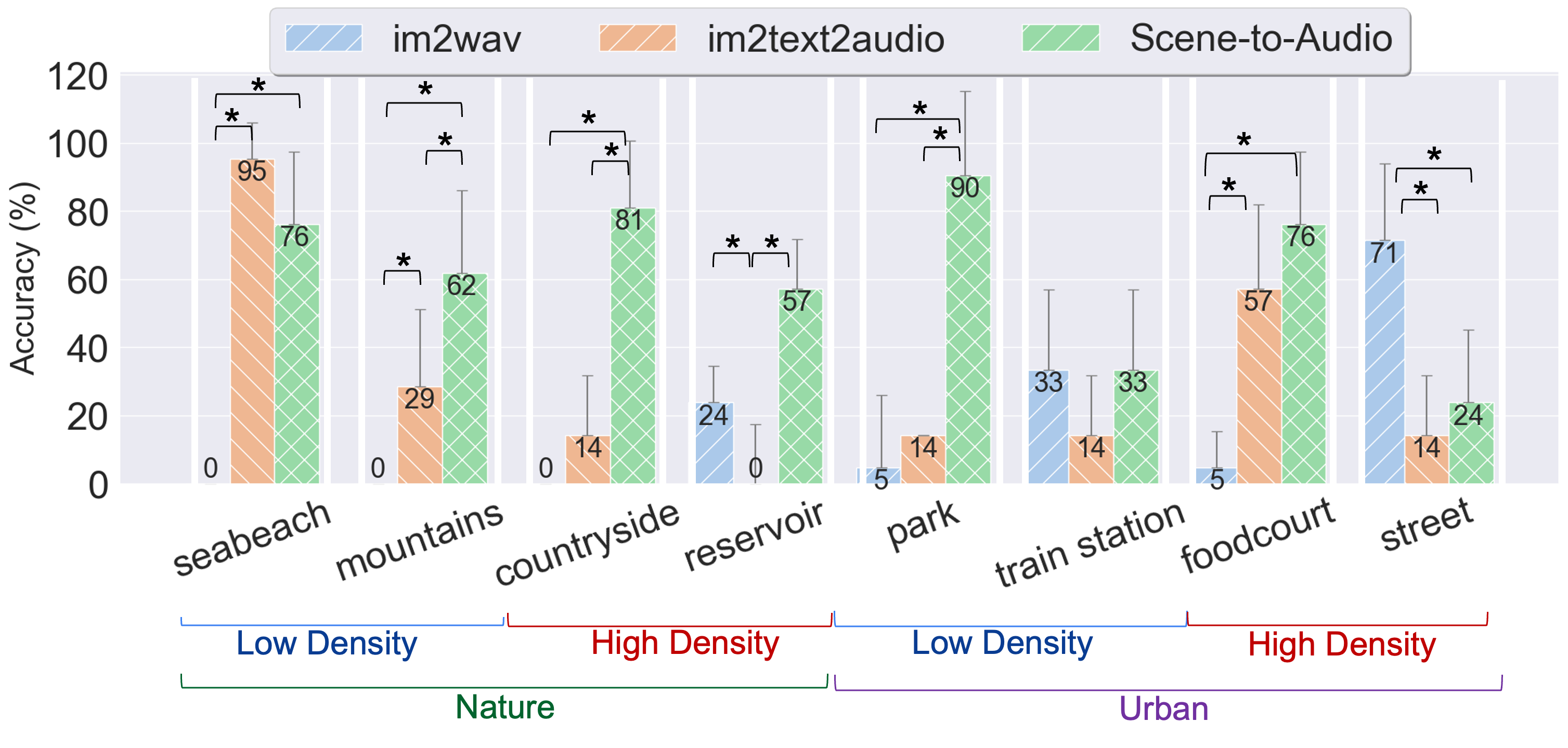}
\vspace{-0.7cm}
\captionof{figure}{Accuracy (in percentage) of image recognition in the listening test, i.e.~accuracy of selecting the image that was used to generate that sound, out of four scene image options. 
Pairs marked with a $*$ indicate statistically significant difference with a p-value $< 0.05$ 
using Wilcoxon signed rank test comparison between the two systems (Please find the exact values in Table \ref{table:comparison}).
} 
\Description{
This bar chart visualizes the accuracy (in percentage) of three different models—im2wav (light blue, with diagonal lines), im2text2audio (orange, with diagonal lines), and Scene2Audio (green, with crosshatch pattern)—across eight distinct environmental scenes. The x-axis represents the environments categorized by density and type: seabeach, mountains, countryside, reservoir (all classified as "Nature"), and park, train station, foodcourt, street (classified as "Urban"). The environments are further divided into low-density and high-density sonic object groups.

The y-axis shows accuracy in percentage, ranging from 0\% to 100\%. Asterisks above bars indicate statistically significant differences between models for a given environment.

Key results include:
- **Seabeach**: im2text2audio has the highest accuracy at 95\%, Scene2Audio follows with 76\%, while im2wav records 0\%. Significant differences exist between Im2Wav and the other two models.
- **Mountains**: Scene2Audio scores highest at 62\%, followed by im2text2audio at 29\%, and im2wav at 0\%. Significant differences are present between scene-to-audio and the other two models.
- **Countryside**: Scene2Audio achieves 81\%, im2text2audio scores 14\%, and im2wav records 0\%. Significant differences are noted between scene-to-audio and the other two models.
- **Reservoir**: Scene2Audio records 57\%, im2text2audio at 24\%, and im2wav at 0\%. Significant differences exist between between im2text2audio and the other two models.
- **Park**: Scene2Audio leads at 90\%, followed by im2text2audio at 14\%, and im2wav at 5\%. Significant differences are observed between scene-to-audio and the other two models.
- **Train station**: Scene2Audio scores 33\%, im2wav 33\%, and im2text2audio 14\%. No statistical significance is noted.
- **Foodcourt**: Scene2Audio achieves 76\%, im2wav 5\%, and im2text2audio 57\%. Significant differences are present between im2wav and the other two models.
- **Street**: Scene2Audio records 24\%, im2wav 71\%, and im2text2audio 14\%. Statistically significant differences are observed between im2wav and the other two models.

Below the chart, the environments are grouped into Nature and Urban categories, with low-density nature scenes (seabeach, mountains) on the left, low-density urban scenes (park, train station) on the right, and high-density nature scenes (reservoir, countryside) on the left, and high-density urban scenes (foodcourt, street).
}
\label{fig:comparison}
\end{minipage}
\hfill
\begin{minipage}{0.34\textwidth}
\centering
\includegraphics[width=\textwidth]{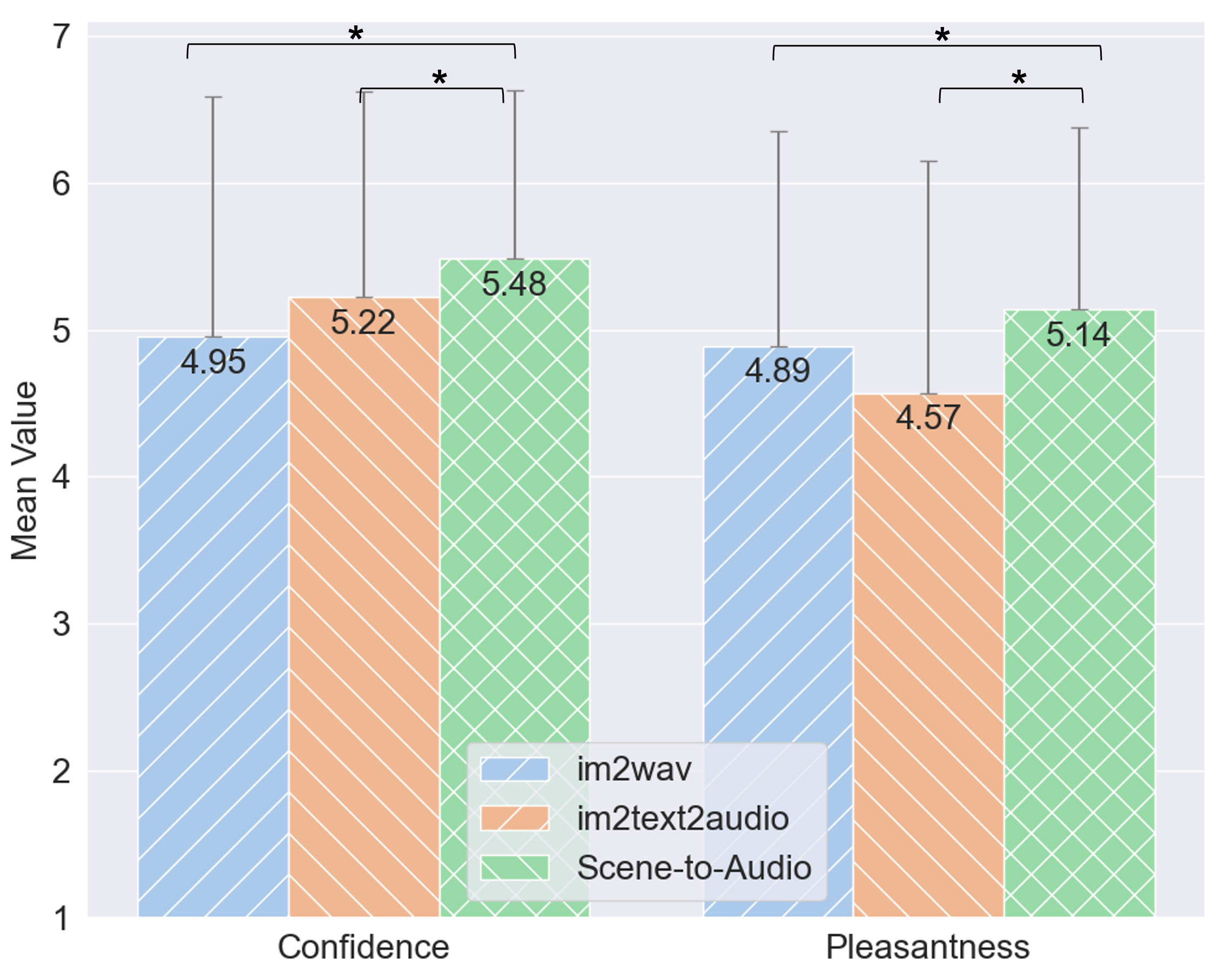}
\captionof{figure}{Confidence of the choice of scene image for a given sound and Pleasantness of the sound on a 7-pt Likert Scale related to the three systems. Pairs marked with a $*$ indicate statistically significant difference with a p-value $< 0.05$ using Wilcoxon signed rank test comparison between the two systems.}
\Description{
This bar chart compares the mean values of 7 point Likert ratings for the two subjective measures, **Confidence** and **Pleasantness**, for three different models: im2wav (light blue, with diagonal lines), im2text2audio (orange, with diagonal lines), and Scene2Audio (green, with crosshatch pattern). The mean values are plotted on the y-axis, ranging from 1 to 7, with error bars representing standard deviations.

- For **Confidence**:
  - im2wav scores a mean of 4.95 ± 1.64.
  - im2text2audio scores a mean of 5.22 ± 1.40.
  - Scene2Audio has the highest mean confidence of 5.48 ± 1.15.
  - Asterisks above the bars indicate statistically significant differences between Scene2Audio and the other two models.

- For **Pleasantness**:
  - im2wav has a mean score of 4.89 ± 1.46.
  - im2text2audio scores a mean of 4.57 ± 1.58.
  - Scene2Audio achieves the highest pleasantness score of 5.14 ± 1.24.
  - Statistically significant differences are noted between Scene2Audio and both im2wav and im2text2audio for pleasantness.

The comparison highlights that the Scene2Audio model scores higher for both confidence and pleasantness, with significant differences from the other two models.
}
\label{fig:confidence_pleasantness}
\end{minipage}
\end{table*}

Subjective evaluation of the confidence in scene image selection for a given sound and pleasantness of the sound (Figure \ref{fig:confidence_pleasantness}) also show similar results. Participants rated the audio generated by our framework significantly higher than baselines for both these metrics, indicating higher perceived confidence in relating a scene to the audio, as well as perceived pleasantness of the sounds (Wilcoxon signed rank test p < 0.05).

Qualitative feedback provides additional insights. In some low-density scenes, performance between methods was similar. For instance, participants remarked that both Im2text2audio and our framework generated similar seabeach sounds. However, in high-density scenes, baseline methods struggled to sonify all salient objects in the scene, leading participants to select distractor images. For example, when listening to the countryside sound generated by Im2wav, a participant stated, \textit{``I hear church bells but no animals, so I chose the one without animals.''} Similarly, for Im2text2audio, another participant mentioned, \textit{``I hear only animals and no church-like sounds.''} However, when evaluating the sound generated by our framework, participants correctly selected the intended image and commented, \textit{``I hear cow moos and a church bell,''} indicating that our method more effectively captured the complexity of the scene.

Our framework did not perform well in urban scenes like \textit{street} and \textit{train station}. 
The traffic light beeping was incorrectly generated and sounded more like objects from the distractor images, such as tram bells. Participants noted that \textit{``Definitely sounds like the beeping of a tram,''} and \textit{``... some bing sound which could be of a train,''}, highlighting the potential for confusion and ambiguity when similar sonic elements are heard in different scenes. On the other hand, the baseline Im2Wav sound only consisted of people murmur in a crowded area that matched better with the correct image than any of the distractor images.

These results highlight the potential 
of the Scene2Audio framework in capturing varied sonic object densities and suggest that it is more adept at conveying the auditory essence of scenes than existing methods. However, the potential ambiguity of non-verbal sounds, especially in urban scenes, indicates the need to explore ways to disambiguate and improve comprehension of the scenes. 
\end{enumerate}

\end{document}